\documentclass{article}

\usepackage{amsmath}
\usepackage{arxiv}
\usepackage[utf8]{inputenc} 
\usepackage[T1]{fontenc}    
\usepackage{hyperref}       
\usepackage{url}            
\usepackage{booktabs}       
\usepackage{amsfonts}       
\usepackage{nicefrac}       
\usepackage{microtype}      
\usepackage{mathtools}
\usepackage{url}
\usepackage{float}
\usepackage{placeins}
\usepackage{setspace}
\usepackage{adjustbox}
\usepackage{lineno}
\usepackage{natbib}

\makeatletter
\newif\ifnobrackets
\renewcommand\@cite[2]{\ifnobrackets\else[\fi{#1\if@tempswa , #2\fi}\ifnobrackets\else]\fi\nobracketsfalse}

\makeatother

\title{Speeding up reactive transport simulations in cement systems by surrogate geochemical modeling: deep neural networks and \textit{k}-nearest neighbors}

\author{
   Eric Laloy\thanks{\normalsize{Corresponding Author, \texttt{eric.laloy@sckcen.be}}} \\
   \And
   Diederik Jacques \\
}

\begin{document}
\maketitle

\begin{abstract}
We accelerate reactive transport (RT) simulation by replacing the geochemical solver in the RT code by a surrogate model or emulator, considering either a trained deep neural network (DNN) or a k-nearest neighbor (kNN) regressor. We focus on 2D leaching of hardened cement paste under diffusive or advective-dispersive transport conditions, a solid solution representation of the calcium silicate hydrates and either 4 or 7 chemical components, and use the HPx (coupled Hydrus-PHREEQC model) reactive transport code as baseline. We find that after training, both our DNN-based and kNN-based codes, HPx$_{\rm py}$-DNN and HPx$_{\rm py}$-kNN, can make satisfactorily to very accurate predictions while providing either a 3 to 9 speedup factor compared to HPx with parallelized geochemical calculations over 4 cores. Benchmarking against single-threaded HPx, these speedup factors become 8 to 33. Overall, HPx$_{\rm py}$-DNN are HPx$_{\rm py}$-kNN are found to achieve a close to optimal speedup when DNN regression and kNN search are performed on a GPU. Importantly, for the more complex 7-components cement system, no emulator that is globally accurate over the full space of possible geochemical conditions could be devised. Instead we therefore build ``local" emulators that are only valid over a relevant fraction of the input parameter space. This space is identified by running a coarse and thus computationally cheap full RT simulation, and subsequently explored by kernel density sampling. Future work will focus on improving accuracy for this type of cement systems.
\end{abstract}

\section{Introduction}
\label{intro}
A novel research direction in reactive transport modeling is to replace the geochemical solver of a reactive transport model (RTM) by a trained machine learning (ML) algorithm to speed up the computations. This is motivated by the fact that the computational cost of a reactive transport (RT) simulation is mostly incurred by the geochemical calculations. Hence simulating geochemical processes typically consumes 80\% to 90\% of the total RT simulation time. Replacing the computationally intensive geochemical solver by a computationally cheap albeit potentially less accurate statistical emulator (or metamodel) is therefore quite attractive. The pioneer study by \citet{Jatnieks2016} presents a short comparison of a range of classical machine learning (ML) methods for the emulation of a geochemical solver within a simple RT scenario. Follow-up advances consider more complex RT problems and various emulation (that is, metamodeling or surrogate modeling) techniques. \citet{Huang2018} use a gridded lookup table for a 1D cement system involving alkalisilica reaction and carbonation. \citet{Leal2020} proposed an on-demand learning algorithm to speed up multiphase chemical equilibrium calculations in reactive transport simulations combining first-order Taylor predictions with and on-demand clustering strategy for faster search operations. Noticeably, the approach by \citet{Leal2020} includes a check on surrogate prediction quality and automatic replacement by a full physics geochemical simulation of those surrogate predictions that are considered to be inaccurate. \citet{Guerillot-Bruyelle2019} replaced the geochemical solver by a trained neural network (NN) for a 3D case study of CO2 storage in geological formation. \citet{Prasianakis2020} used NNs to either replace the porosity-permeability relationship or predict saturation index from the concentration of two chemical components in a reactive transport code. Lastly, in a similar spirit of \citet{Leal2020}, \citet{Delucia-Kuhn2021} present a strategy in which at runtime, full physics geochemical simulations are performed only if surrogate predictions are deemed implausible. 

In this work, we further advance the topic by investigating the replacement of the geochemical solver by a machine learning (or data-driven) surrogate for a series of 2D reactive transport problems involving cement degradation. We consider two types of transport conditions: advection and diffusion, and replace the PHREEQC-3 \citep{Parkhurst-Appelo2013} geochemical simulator in the HPx reactive transport code \citep{Jacques2018} either by a kNN nearest neighbor algorithm or by a deep neural network (DNN) \citep[see, e.g.,][]{Goodfellow2016} trained beforehand. In terms of geochemistry and associated dimensionality of the surrogate model, we consider non-trivial cement systems with C-S-H solid solutions that involve both a 2-input (Ca,Si) - 4-output (Ca,Si,O,H) setup and a 5-input (Ca,Si,Al,C,S) - 7-output (Ca,Si,Al,C,S,O,H) setup. To overcome the observed drawbacks associated with using a purely random training set to build the considered emulators, we propose an approach where geochemical knowledge obtained from a small (and thus computationally cheap) physics-based RT simulation is used to guide the construction of the training set. This process ranges from simply extracting lower and upper bounds for the input space to identifying which portions of the input space should be preferentially represented in the training set. In addition, our coupling between the transport model and the geochemical surrogate is very flexible as we plug a call to Python within the mainly C/C++ written HPx code. Doing so, we can use any emulation method that comes with a Python interface. This is convenient as a large fraction of the open-source ML tools and libraries are Python-based. 

The remainder of this paper is organized as follows. Section \ref{methods} describes the considered reactive transport problems together with the used kNN and DNN algorithms and their implementation within HPx. Section \ref{results} then details our results in terms of both speedup and simulation accuracy. This is followed by a discussion of the current limitations and potential improvements of our approach in section \ref{discussion}, before section \ref{conclusion} provides a conclusion.

\section{Methods}
\label{methods}
\subsection{Reactive transport problems}
\label{rt_prob}

Our reactive transport setups calculate leaching of hardened cement paste under diffusive or advective-dispersive transport conditions. Two cement systems are considered: (i) a relatively simple Ca-Si-O-H system, and (ii) a more representative Al-C-Ca-S-Si-H-O system. Both systems consider portlandite and a solid solution representation of the calcium silicate hydrates (C-S-H) using the CSHQ model by \citet{Kulik2011}. The second system also contains calcite, straetlingite, monocarboaluminate and ettringite. Thermodynamic properties for the aqueous and solid phases are taken from CEMDATA18 \citep[][]{Lothenbach2018}. For both cement systems, the problem is defined with a small amount of hardened cement paste to obtain leaching fronts in affordable calculation times. The initial condition is obtained by hydrating 10 g/dm$^{3}$ of cement clinkers with the composition $\rm{\left[CaO, SiO_2, CO_2, Al_2O_3, SO_3\right] = \left[1.11, 0.314, 0.0477, 0.432, 0.0375\right]}$ mol/100 g with a water-cement ratio of 0.5 (for cement system 1, only CaO and SiO$_2$ are used). A porosity of 0.5 is considered. The initial aqueous phase is in equilibrium with the hardened cement paste (Table \ref{table1}). The two-dimensional flow and transport field measures 3 $\times$ 3 cm$^2$. All boundaries are closed, except a 1 cm wide open boundary at the top right and bottom left sides. In case of advective-dispersive transport and for solving the Richards equation for water flow with HPx, the boundary conditions at the open parts are set by defining a constant pressure head of 30 cm at the top and 0 cm at the bottom. For solute transport, a constant concentration (first type) and a constant concentration flux (third type) boundary condition  are assumed for the diffusive and advective-dispersive transport conditions, respectively. Dilute water (with 1 $\mu$mol of each of the chemical components) is entering the system at the top boundary of the system. By considering equilibrium, dissolution/leaching of the cement hydrates is simulated considering the minerals listed in Table \ref{table1}. 

We test with 61 $\times$ 61 and 121 $\times$ 121 grids for both transport conditions. This choice is dictated by computational expense and our available hardware: 4 CPUs Intel i7 2.70 Ghz. To keep the computations required to obtain the benchmark simulations tractable, the considered simulation time period varies between 2 and 6 years. Computational time is detailed for each experiment later on.

\begin{table}[h!]
	\caption{Initial mineralogical composition for the 2 cement systems (calculated with CEMDATA18 \citep{Lothenbach2018}). Units are mol/dm$^3$.}
	\begin{center}
		\begin{tabular}{cccc}%
			\hline
			& & & \\
			Mineral & End members & System 1 & System 2\\
			Portlandite & & 5.9 $\times$ 10$^{-2}$ & 3.8 $\times$ 10$^{-2}$\\
			Monocarbonate & & & 3.1 $\times$ 10$^{-3}$ \\
			Ettringite & $NA$* & & 1.2 $\times$ 10$^{-3}$ \\
			Straetlingite & & & 0 \\
			Calcite & & & 1.7 $\times$ 10$^{-3}$\\
			C-S-H (ideal solid solution)  & & & \\
			& CSHQ-JenD & 1.7 $\times$ 10$^{-2}$ & 1.7 $\times$ 10$^{-2}$ \\
			& CSHQ-JenH & 1.1 $\times$ 10$^{-2}$ & 1.1 $\times$ 10$^{-2}$ \\
			& CSHQ-TobD & 1.3 $\times$ 10$^{-2}$ & 1.3 $\times$ 10$^{-2}$ \\
			& CSHQ-TobH & 5.6 $\times$ 10$^{-4}$ & 5.6 $\times$ 10$^{-4}$ \\
			\hline
			\multicolumn{4}{l}{*Not Applicable.}  
		\end{tabular}
	\end{center}
	\label{table1}
\end{table}

\FloatBarrier

\subsection{Emulation strategy and implementation}
\label{rt_emul}

The coupled reactive transport model for leaching of hardened cement paste is implemented in the HPx code that couples Hydrus \citep{Simunek2013}  with PHREEQC \citep{Parkhurst-Appelo2013} using a sequential non-iterative approach \citep{Jacques2018}. Transport is calculated for each chemical component, i.e. in terms of total aqueous concentration of the given element. After the independent transport calculations in each time step, geochemical calculations with PHREEQC are done for each grid node to calculate the equilibrium solid phase and aqueous composition. As stated earlier, we test replacing the PHREEQC geochemical solver of HPx by a trained nonlinear regressor which we refer to as an emulator (also called commonly metamodel, surrogate model or proxy model). For each time step and grid node of a given reactive transport simulation, we emulate the components’ aqueous concentrations from the total components’ amounts and then re-calculate the components’ solid amounts by subtracting for each component the new aqueous concentration from the total amount. The total mass before and after a reactive transport step in a single cell is therefore fully conserved, although because of the emulation error it can be wrongly distributed between solid and aqueous phases. 

Our emulators are Python-based and a call to the Python language is introduced within the C/C++ written HPx code. We refer to the resulting HPx variant as HPx$_{\rm py}$. When benchmarking against HPx, we consider both the open-mp version where the PHREEQC calculations are parallelized over the physical cores of the computer (in our case 4 cores), that we refer to as four-core HPx or HPx$_{\rm{4C}}$, and single-threaded or single-core HPx that we call  HPx$_{\rm{1C}}$. Regarding terminology, we refer to the HPx-simulated data as ``original" data and the simulated data by HPx$_{\rm py}$-DNN and HPx$_{\rm py}$-kNN as emulated data. Importantly, for the problems considered herein transport calculations roughly represent 10 \% to 20 \% of the total reactive transport simulation time with HPx$_{\rm{4C}}$. Defining speedup as ``number of times faster", this means that the corresponding maximum possible speedup, which would be obtained if the PHREEQC-based geochemical computations would incur no computational cost at all, ranges between 5 and 10. If HPx$_{\rm{1C}}$ is used, that is, if all of reactive transport computations are achieved on a single thread, then the runtime fraction associated with transport decreases to between approximately 3 \% and 5.5 \%, while the associated maximum possible speedup increases to between 18 and 33.

The emulation techniques investigated in this study are k-nearest neighbors \citep[kNN, e.g.,][]{Hastie2009} and deep neural networks \citep[DNN, e.g.,][]{Goodfellow2016}. The main reason for this choice is that these techniques are very fast while a large prediction speed is needed for the emulator to compete against geochemical solvers such as PHREEQC, which for the considered two cement systems and hardware performs either about 670 (system 1) or 210 (system 2) calculations per second on a single thread  (Intel\textsuperscript{\textregistered} i7 2.70GHz CPU). Furthermore we deal herein with multi-output regression and both kNN and DNN attempt to honor, in their own distinctive ways, the relationships between the different output targets. That makes kNN and DNN attractive compared to emulation approaches that require training a separate regressor for each output target, which (i) does not leverage any possible relation between targets and (ii) is likely to be slower than multi-output emulators. As further detailed later on, for our cement system 1 (2 inputs  - 4 outputs) and a training base of 400,000 examples, single-threaded kNN is found to be about 300 times faster than single-threaded PHREEQC for performing 10,000 calculations. This speedup becomes as large as 4000 for our trained DNN when ran on a NVIDIA Quadro M2000M GPU. For our cement system 2 (5 inputs - 7 outputs), our trained GPU-based DNN remains approximately 1000 (3000) times faster than single-threaded PHREEQC when ran on a NVIDIA Quadro M2000M (NVIDIA Quadro P6000) GPU. For this second problem, standard (and single-threaded) kNN becomes prohibitively slow compared to HPx$_{\rm{4C}}$ and we thus rely on an approximate GPU-compatible kNN algorithm that, as detailed later, we run on NVIDIA Quadro P6000 GPU. For a training base of 1,000,000 samples, this GPU-powered approximate kNN algorithm is also about 3000 times faster than single-threaded PHREEQC for producing 10,000 predictions at once. Importantly, for all of these comparisons PHREEQC is run in batch mode and is thus initialized only once.

The kNN technique basically finds a number of similar instances to a presented example within a training base using a given distance measure, and then interpolate between them. Our used kNN regressor for the considered first cement system is the kNN regressor implemented in the Python scikit-learn toolbox \citep[][]{sklearn}, using the default automatic selection between the ``ball tree" and ``k-d tree" methods for exact nearest neighbor search \citep[see,][for details]{sklearn}. We search for the 5 closest neighbors with respect to the euclidean distance and perform an inverse-distance weighted interpolation. Furthermore, we run our scikit-learn kNN on a single-thread since we observed a drop in computation speed using the multi-threading option, when called from within our Windows-based HPx$_{\rm py}$ framework. As mentioned above, for the considered second cement system the scikit-learn kNN regression approach becomes too slow compared to  HP$_{\rm{4C}}$. Therefore, we implemented our own kNN regression based on the GPU-powered FAISS package for kNN search, using an approximate search method \citep[see][for algorithmic details]{faiss2017}.

Neural networks basically define the (possibly complex) relationships existing between input, $\textbf{x}$, and output, $\textbf{y}$, data vectors by using combinations of computational units that are called neurons. A neuron is an operator of the form:
\begin{equation}
h\left(\textbf{x}\right) =f\left(\langle \textbf{x}, \textbf{w} \rangle+ b \right),
\label{dnn1}
\end{equation}
where $h\left(\cdot \right)$ is the scalar output of the neuron, $f\left(\cdot \right)$ is a nonlinear function that is called the ``activation function", $\langle\cdot,\cdot\rangle$ signifies the scalar product, $\textbf{w} = \left[w_1, \cdots, w_N\right]$ is a set of weights of same dimension, $N$, as $\textbf{x}$ and $b$ represents the bias associated with the neuron. For a given task, the values for $\textbf{w}$ and $b$ associated with each neuron must be optimized or ``learned" such that the resulting neural network performs as well as possible. When $f\left(\cdot \right)$ is differentiable, $\textbf{w}$ and $b$ can be learned by gradient descent. Common forms of $f\left(\cdot \right)$ include the rectified linear unit (ReLU), sigmoid function and hyperbolic tangent function.

When there is no directed loops or cycles across neurons or combinations thereof, the network is said to be feedforward (FFN). In the FFN architecture, the neurons are organized in layers. A standard layer is given by
\begin{equation}
\textbf{h}\left(\textbf{x}\right)=f\left(\textbf{W}\textbf{x} + \textbf{b} \right),
\label{dnn2}
\end{equation}
where $\textbf{W}$ and $\textbf{b}$ are now a matrix of weights and a vector of biases, respectively. The name multilayer perceptron (MLP) designates a FFN with more than one layer that is fully connected (FC), that is, where every neuron of a given layer is connected to all neurons of the next layer. A most typical network is the 2-layer MLP, which consists of two layers with the outputs of the first-layer neurons becoming inputs to the second-layer neurons
\begin{equation}
\textbf{y}=\textbf{g}\left[\textbf{h}\left(\textbf{x}\right)\right]  \equiv f_2\left[\textbf{W}_2\cdot f_1\left(\textbf{W}_1\textbf{x} + \textbf{b}_1 \right) + \textbf{b}_2 \right],
\label{dnn3}
\end{equation}
where $\textbf{g}\left(\cdot\right)$ and $\textbf{h}\left(\cdot\right)$ are referred to as output layer and hidden layer, respectively.

In theory, the two-layer MLP described in equation (\ref{dnn3}) is a universal approximator as it can approximate any underlying process between $\textbf{y}$ and $\textbf{x}$ \citep{Cybenko1989,Hornik1991}. However, this only works if the dimension of $\textbf{h}\left(\cdot\right)$ is (potentially many orders of magnitudes) larger than that of the input $\textbf{x}$, thereby making learning practically infeasible and the two-layer MLP approximator useless for large $N$ (typically $N \geq$ 5 - 10). Practitioners have found that it is much more efficient to use many hidden layers rather than increasing the size of a single hidden layer \citep[e.g.,][]{Goodfellow2016}. When a FFN/MLP has more than one hidden layer it is considered to be deep. Nevertheless, current deep networks are not necessarily purely FFN but may mix different aspects of FFN, such as convolutional neural networks (CNN) and recurrent neural networks \citep[RNN, see, e.g.,][]{Goodfellow2016}.

Our selected DNN networks are detailed in Appendix. They basically consist of a 6-layers FC neural network with scaled exponential linear units (SELUs) as activation functions. SELUs have been showed to outperform the widely used rectified linear units (RELUs) for training fully connected networks \citep[][]{Klambauer2017}. The size of our FC layers progressively increases from input dimensionality to a maximum size of either 128 (cement sytem 1) or 512 neurons (cement system 2), and then decreases stepwise towards the output dimensionality. Our DNNs are implemented within the pytorch framework \citep{pytorch2017} and training is performed by stochastic gradient descent with the Adam algorithm \citep[][]{Kingma-Ba2015}. 

All GPU calculations were performed on a NVIDIA Quadro M2000M GPU for cement system 1, and on a more recent NVIDIA Quadro P6000 GPU for cement system 2 (mainly because the Windows-based FAISS kNN GPU implementation used for cement system 2 is not compatible with a GPU as old as the  NVIDIA Quadro M2000M). Running DNNs and kNN on a GPU (or more if available) is significantly faster than running on CPUs.

\subsection{Metrics for training performance assessment}
\label{metrics}
We resort to two metrics to assess the training quality of each emulator using an independent test set of 10,000 samples, $\left[\textbf{X}^*,\textbf{Y}^*\right]$, that is therefore not used for training. The $Q_2$ coefficient (also often referred to as coefficient of determination)is given by
\begin{equation}
Q_2 = 1-\frac{\sum_{i = 1}^{n^*}\sum_{j = 1}^{d_y}\left(y^*_{i,j} - y^*_{s,i,j}\right)^2}{\sum_{i = 1}^{n^*}\sum_{j = 1}^{d_y}\left(y^*_{i,j} - \overline{\textbf{Y}^*}\right)^2},
\label{q2}
\end{equation}
where $\textbf{Y}^*_s$ is a $n^* \times d_y$ array of simulated outputs and $\overline{\textbf{Y}^*}$ denotes the mean of $\textbf{Y}^*$. Furthermore, the root-mean-square-error (RMSE) is defined as
\begin{equation}
{\rm RMSE} =\sqrt{\frac{\sum_{i = 1}^{n^*}\sum_{j = 1}^{d_y}\left(y^*_{i,j} - y^*_{s,i,j}\right)^2}{n^*d_y}}.
\label{rmse}
\end{equation}

\section{Results}
\label{results}

\subsection{Ca-Si Problem}
For this first cement system, the emulation problem consists of predicting at each time step of the RT simulation the (output) Ca, Si, H and O aqueous concentrations (mol/kg of water or mol/kgw) from the (input) total amounts of Ca and Si (mol). 

\subsubsection{Training the emulators}
\label{train_res1}
Here the kNN and DNN emulators are firstly trained using a set of 400,000 test examples for both. This training set is obtained by randomly sampling the two-dimensional input space by latin hypercube sampling (LHS) between $\left[0,0\right]$ and $\rm{\left[Ca^{tot}_{max},Si^{tot}_{max}\right]}$, and running PHREEQC for each input sample, $\rm{\textbf{x}_i = \left[Ca^{tot}_i, Si^{tot}_i\right]}$ to get the corresponding output vectors, $\rm{\textbf{y}_i = \left[Ca^{conc}_i,Si^{conc}_i,H^{conc}_i,O^{conc}_i\right]}$. The upper bounds, $\rm{Ca^{tot}_{max}}$ and $\rm{Si^{tot}_{max}}$ are defined based on a cheap full RT simulation with advective-dispersive transport using a small 1D domain of 51 nodes. It is worth noting that the total amounts of $\textbf{x}_i$, corresponding to the PHREEQC-simulated concentrations, $\textbf{y}_i$, have to be corrected for the different amount of water between the training set and the transport simulations. Doing so, it turns out that about 20 \% of the post-corrected $x_i$ values exceed their pre-defined upper bounds and these excessively large values need to be filtered out. Creating the 400,000 training examples thus required about 500,000 PHREEQC runs. As stated earlier, for this problem single-threaded PHREEQC performs about 670 geochemical calculations per second on our used Intel\textsuperscript{\textregistered} i7 CPU.

With respect to kNN, the tuning parameters are the number of neighbors, $k$, the type of distance measure, and the interpolation technique. We simply used the default settings: $k = 5$, euclidean distance and inverse-distance interpolation. Regarding training of the DNN, the 400,000 sample were split between the training set itself (90 \% of the data) and a validation set (10 \% of the data). The latter serves to monitor the evolution of the selected mean squared error loss function on samples that are not used for training, thereby detecting potential overfitting. If the validation loss stops decreasing before the fixed number of epochs has been completed, then  training is stopped. Importantly, the emulation is achieved in log-space for both the input, $X$ and output $Y$, domains. This because total amounts and concentrations of the involved components typically cover many orders of magnitudes (up to 10 orders or more). Using a DNN also requires some form of data normalization or standardization. Here both the $\rm{log\left(\textbf{x}_i\right)}$ and $\rm{log\left(\textbf{y}_i\right)}$ vectors are standardized around 0 with standard deviation of 1.

\begin{figure}[h!]
	\noindent\hspace{0cm}\includegraphics[width=35pc]{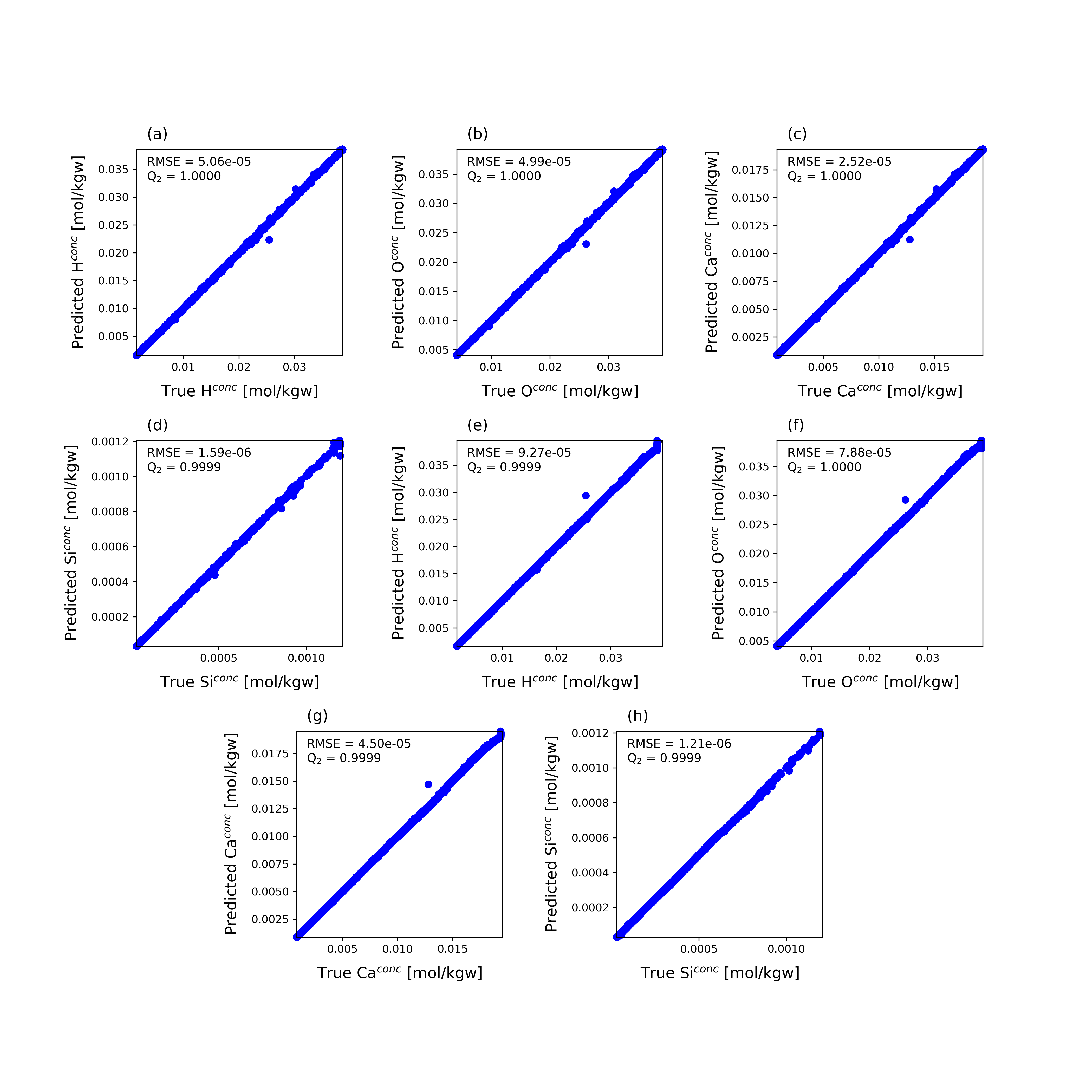}
	\caption{1-1 plots of the kNN (subfigures (a) - (d)) and DNN (subfigures (e) - (h)) emulators' performance obtained for system 1 when the kNN training base contains 400,000 samples and the DNN is trained using the same 400,000 samples. Here ``true" means the original PHREEQC-simulated data and ``predicted" denotes the emulated (that is, kNN-simulated and DNN-simulated) data. Hence, the $x$-axis and $y$-axis present the original and emulated 10,000 independent test data points, respectively. The RMSE and $Q_2$ coefficient denote the root-mean-square-error and coefficient of determination in testing mode, respectively, between the original and emulated 10,000 test data points.}
	\label{fig1}
\end{figure}

Figure \ref{fig1} illustrates the trained emulators' performance for geochemical predictions using an independent test set that comprises 10,000 test examples. Both kNN (Figures \ref{fig1}a - d) and DNN (Figures \ref{fig1}e - h) appear to be rather accurate . DNN also shows a slight degradation for the larger concentration values  (Figures \ref{fig1}e - h). The latter is likely due to the combination of a small proportion of large concentration values in the training set with the log-transformation that implicitly pushes the DNN to try harder to fit the smaller concentrations during training. Regarding speedup and as written earlier, for this setup the single-threaded kNN method is 300 times faster than single-threaded PHREEQC for predicting the 10,000 concentration vectors all at once. The computational savings allowed by the DNN emulator when ran on our GPU is higher with a speedup as large as 4000 for predicting the same 10,000 concentration vectors all at once. 

\begin{figure}[h!]
	\noindent\hspace{0cm}\includegraphics[width=35pc]{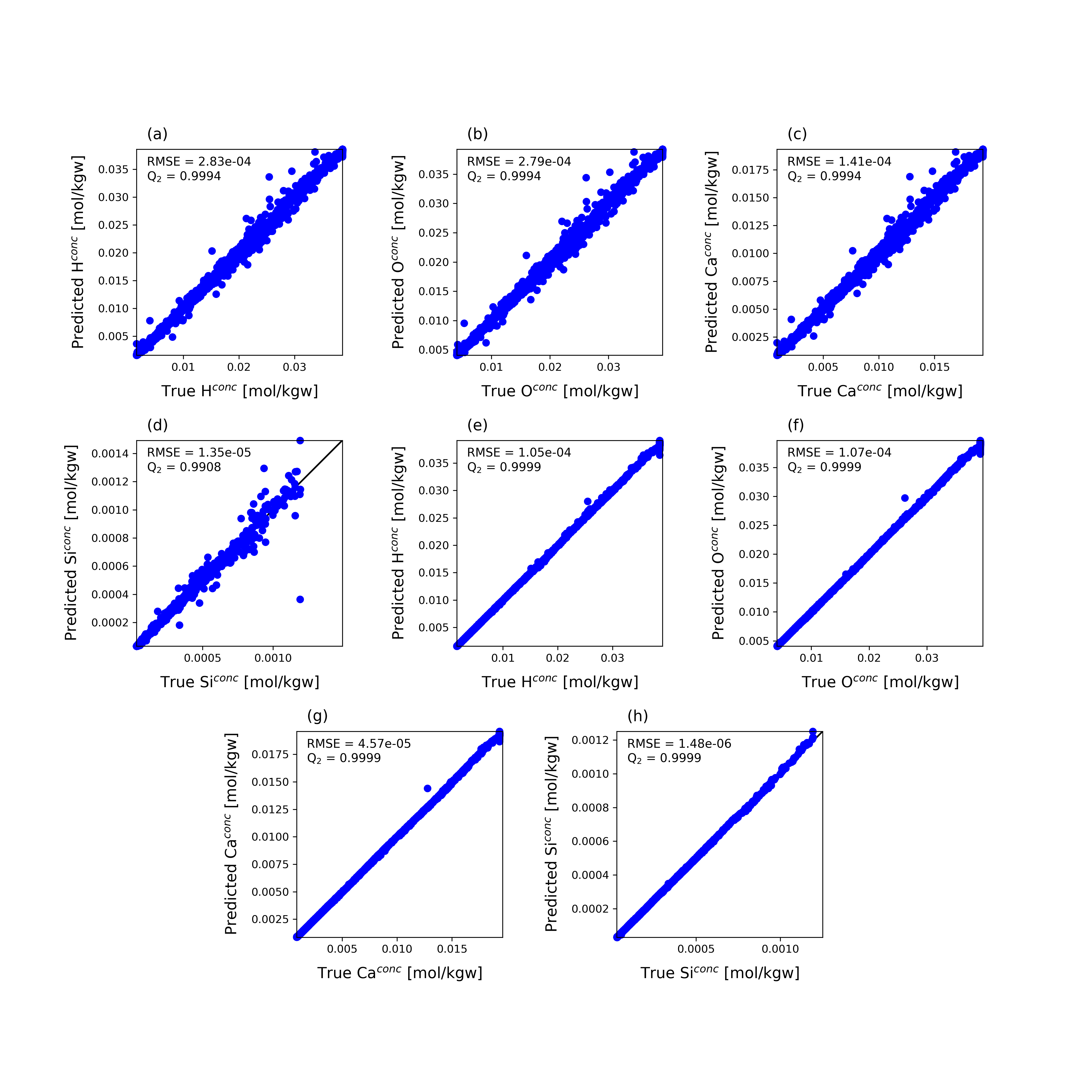}
	\caption{1-1 plots of the kNN (subfigures (a) - (d)) and DNN (subfigures (e) - (h)) emulators' performance obtained for system 1 when the kNN training base contains 10,000 samples and the DNN is trained using the same 10,000 samples. Here ``true" means the original PHREEQC-simulated data and ``predicted" denotes the emulated (that is, kNN-simulated and DNN-simulated) data. Hence, the $x$-axis and $y$-axis present the original and emulated 10,000 independent test data points, respectively. The RMSE and $Q_2$ coefficient denote the root-mean-square-error and coefficient of determination in testing mode, respectively, between the original and emulated 10,000 test data points.}
	\label{fig2}
\end{figure}

To test the sensitivity of the emulators' performance to the training set size, training was also performed using reduced training sets comprising 100,000, 10,000, 1000 and 100 samples, respectively. It seen that the DNN performance achieved  when using 10,000 training samples is virtually the same as that obtained when using 400,000 training samples (Figure \ref{fig2} and Table \ref{table2}). It is only for training sets smaller than 1000 samples that the DNN performance starts to degrade significantly (Table \ref{table2}). In contrast, the behavior of kNN appears to markedly decrease as the training set gets smaller (Figure \ref{fig2} and Table \ref{table1}).

\begin{table}[h!]
	\caption{Performance of the DNN and kNN emulators for cement system 1 and different training set sizes. For brevity, only the results for Ca$^{conc}$ and Si$^{conc}$ are shown. The units are mol per kg of water (mol/kgw). ML refers to the type of emulator, TR signifies the size of the training set, and the RMSE and $Q_2$ coefficient denote the root-mean-square-error and coefficient of determination in testing mode, respectively, between the original and emulated 10,000 test data points.}
	\begin{center}
		\begin{tabular}{cccccccc}%
			\hline
			& & & & & \\
			ML & TR & RMSE - Ca$^{conc}$ & $Q_2$ - Ca$^{conc}$ & RMSE - Si$^{conc}$ & $Q_2$ - Si$^{conc}$\\
			DNN & 4 $\times$ 10$^{5}$ & 4.50 $\times$ 10$^{-5}$ & 0.9999 & 1.21 $\times$ 10$^{-6}$ & 0.9999\\
			DNN & 1 $\times$ 10$^{5}$ & 5.06 $\times$ 10$^{-5}$ & 0.9999 & 1.30 $\times$ 10$^{-6}$ & 0.9999\\
			DNN & 1 $\times$ 10$^{4}$ & 4.57 $\times$ 10$^{-5}$ & 0.9999 & 1.48 $\times$ 10$^{-6}$ & 0.9999\\
			DNN & 1 $\times$ 10$^{3}$ & 8.02 $\times$ 10$^{-5}$ & 0.9998 & 3.31 $\times$ 10$^{-6}$ & 0.9994\\
			DNN & 1 $\times$ 10$^{2}$ & 47.1 $\times$ 10$^{-5}$ & 0.9935 & 9.02 $\times$ 10$^{-6}$ & 0.9959\\
			kNN & 4 $\times$ 10$^{5}$ & 2.52 $\times$ 10$^{-5}$ & 1.0000 & 1.59 $\times$ 10$^{-6}$ & 0.9999\\
			kNN & 1 $\times$ 10$^{5}$ & 6.27 $\times$ 10$^{-5}$ & 0.9999 & 4.20 $\times$ 10$^{-6}$ & 0.9991\\
			kNN & 1 $\times$ 10$^{4}$ & 14.1 $\times$ 10$^{-5}$ & 0.9994 & 13.5 $\times$ 10$^{-6}$ & 0.9908\\
			kNN & 1 $\times$ 10$^{3}$ & 43.9 $\times$ 10$^{-5}$ & 0.9944 & 74.6 $\times$ 10$^{-6}$ & 0.7180\\
			kNN & 1 $\times$ 10$^{2}$  & 96.4 $\times$ 10$^{-5}$ & 0.9729 & 55.0 $\times$ 10$^{-6}$ & 0.8468\\
		\end{tabular}
	\end{center}
	\label{table2}
\end{table}

\FloatBarrier

\subsubsection{Reactive transport simulation}
\label{prob12Dres}

\begin{figure}[h!]
	\noindent\hspace{-1cm}\includegraphics[width=45pc]{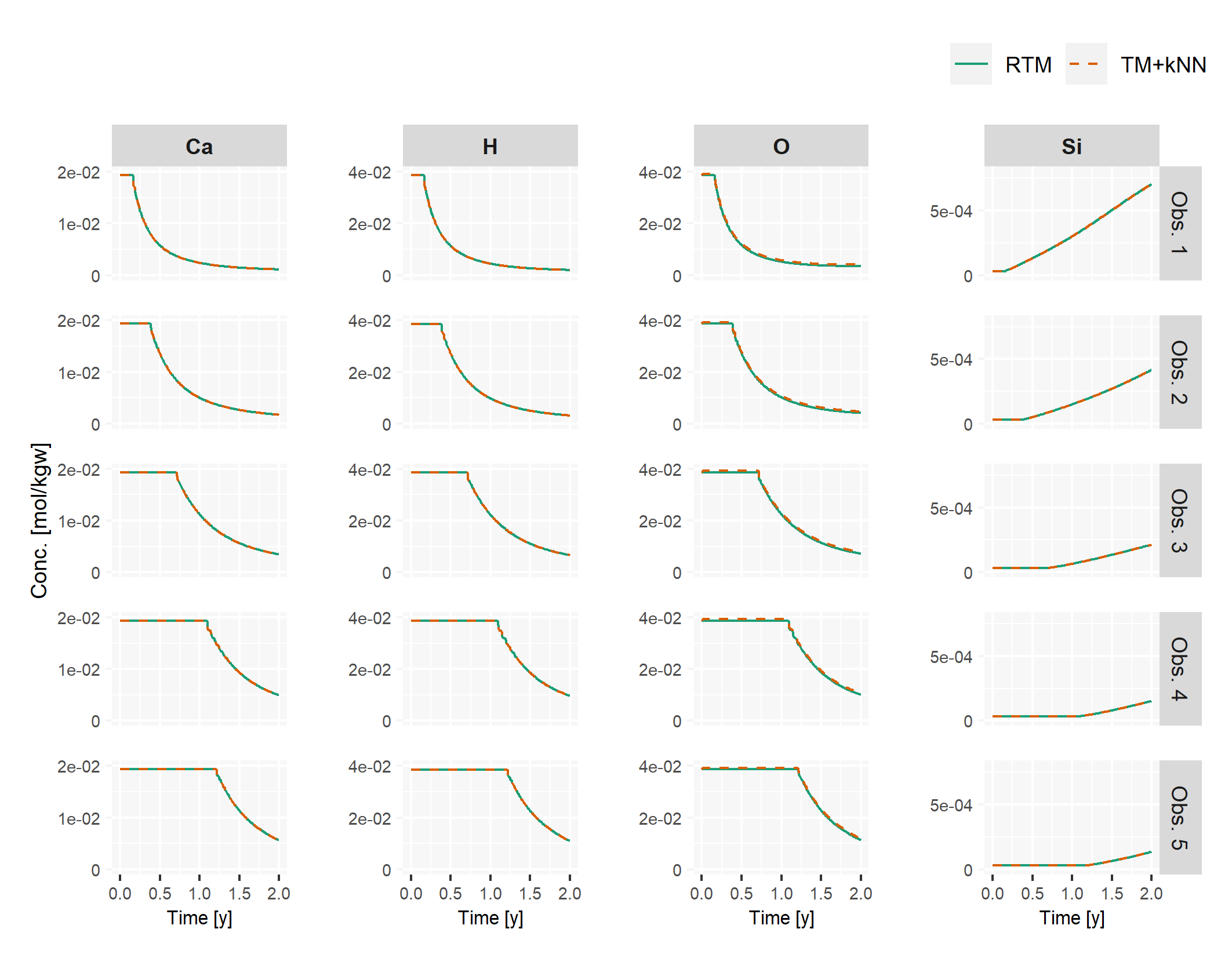}
	\caption{Time series of original (RTM, solid green lines) and HPx$_{\rm py}$-kNN emulated (TM+kNN, dashed orange lines) concentrations (mol/kg) of Ca, H, O and Si at selected observations points for cement system 1 and advective-dispersive transport. Obs. 1 - 5 denote the selected observation points, with the following $\left[x,y\right]$ locations (in cm). Obs. 1:  $\left[0.5, 2.5\right]$, Obs. 2: $\left[1, 2\right]$, Obs. 3: $\left[2, 2\right]$, Obs. 4: $\left[1, 1\right]$, Obs. 5: $\left[2, 1\right]$. The results for the 121 $\times$ 121 grid size are rather similar.}
	\label{fig3}
\end{figure}

\begin{figure}[h!]
	\noindent\hspace{-1cm}\includegraphics[width=45pc]{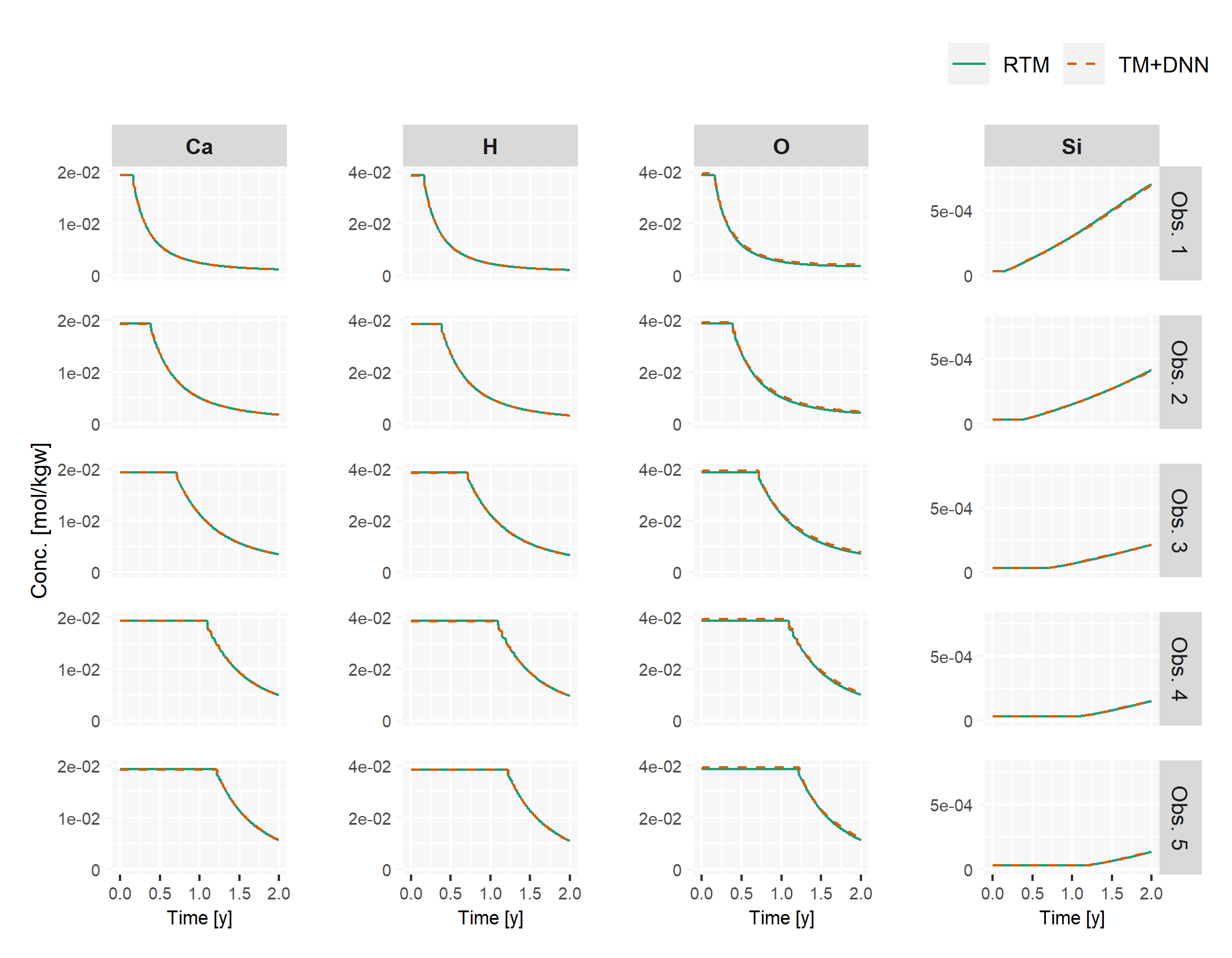}
	\caption{Time series of original (RTM, solid green lines) and HPx$_{\rm py}$-DNN emulated (TM+DNN, dashed orange lines) concentrations (mol/kg) of Ca, H, O and Si at selected observations points for cement system 1 and advective-dispersive transport. Obs. 1 - 5 denote the selected observation points, with the following $\left[x,y\right]$ locations (in cm). Obs. 1:  $\left[0.5, 2.5\right]$, Obs. 2: $\left[1, 2\right]$, Obs. 3: $\left[2, 2\right]$, Obs. 4: $\left[1, 1\right]$, Obs. 5: $\left[2, 1\right]$. The results for the 121 $\times$ 121 grid size are rather similar.}
	\label{fig4}
\end{figure}

\begin{figure}[h!]
	\noindent\hspace{-1cm}\includegraphics[width=45pc]{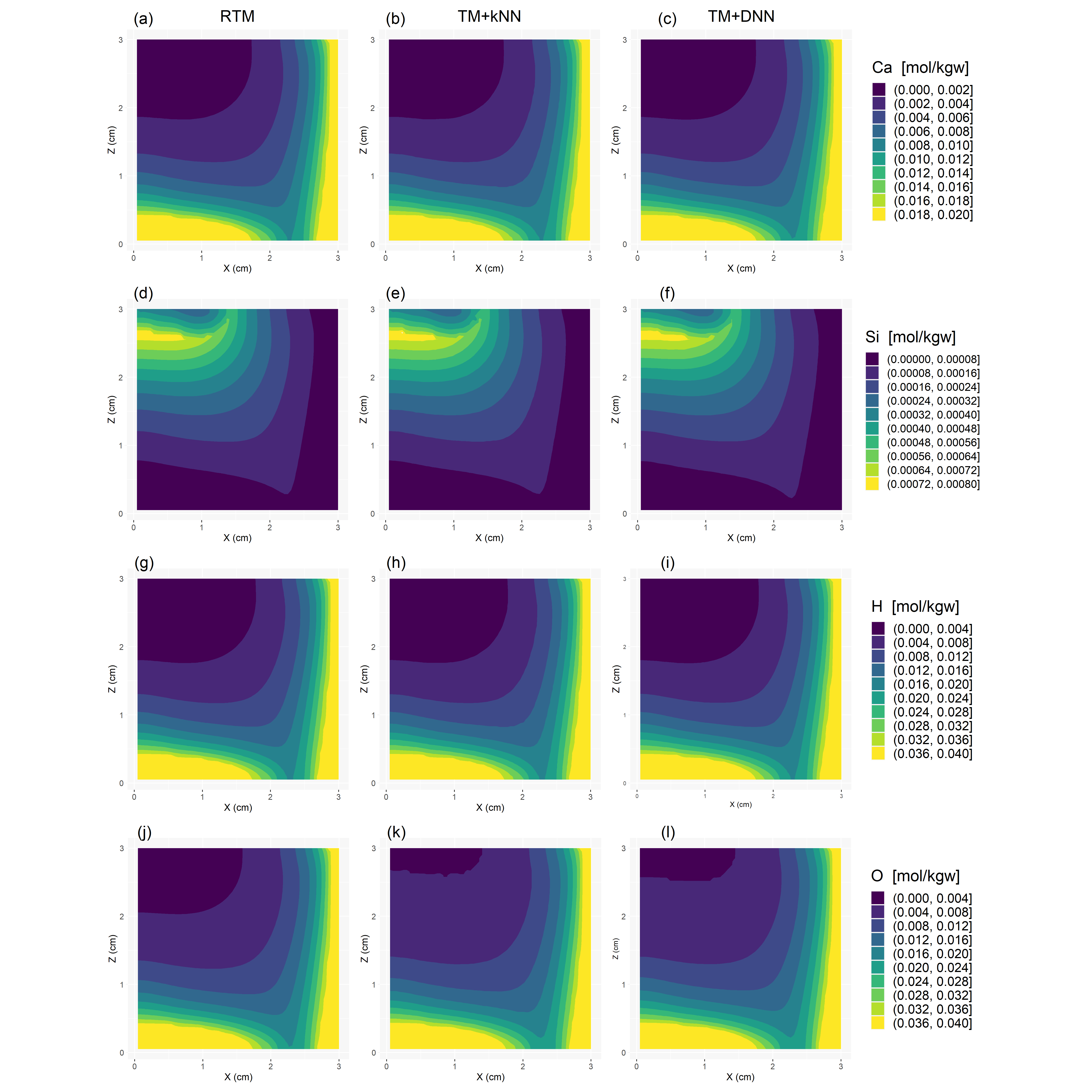}
	\caption{2D concentration profiles obtained for cement system 1 at the end of the 2-year simulation performed for the advection-dispersion case. RTM means the original HP$_{\rm{4C}}$ model, TM+kNN denotes the Hydrus transport model coupled with our kNN geochemical emulator (HPx$_{\rm py}$-kNN) and TM+DNN signifies the Hydrus transport model coupled with our DNN geochemical emulator (HPx$_{\rm py}$-DNN). The first to fourth row present profiles for $Ca^{\rm{conc}}$, $Si^{\rm{conc}}$, $H^{\rm{conc}}$, and $O^{\rm{conc}}$, respectively. The considered grid size is 61 $\times$ 61. The results for the 121 $\times$ 121 grid are rather similar.}
	\label{fig5}
\end{figure}

\begin{figure}[h!]
	\noindent\hspace{-1cm}\includegraphics[width=45pc]{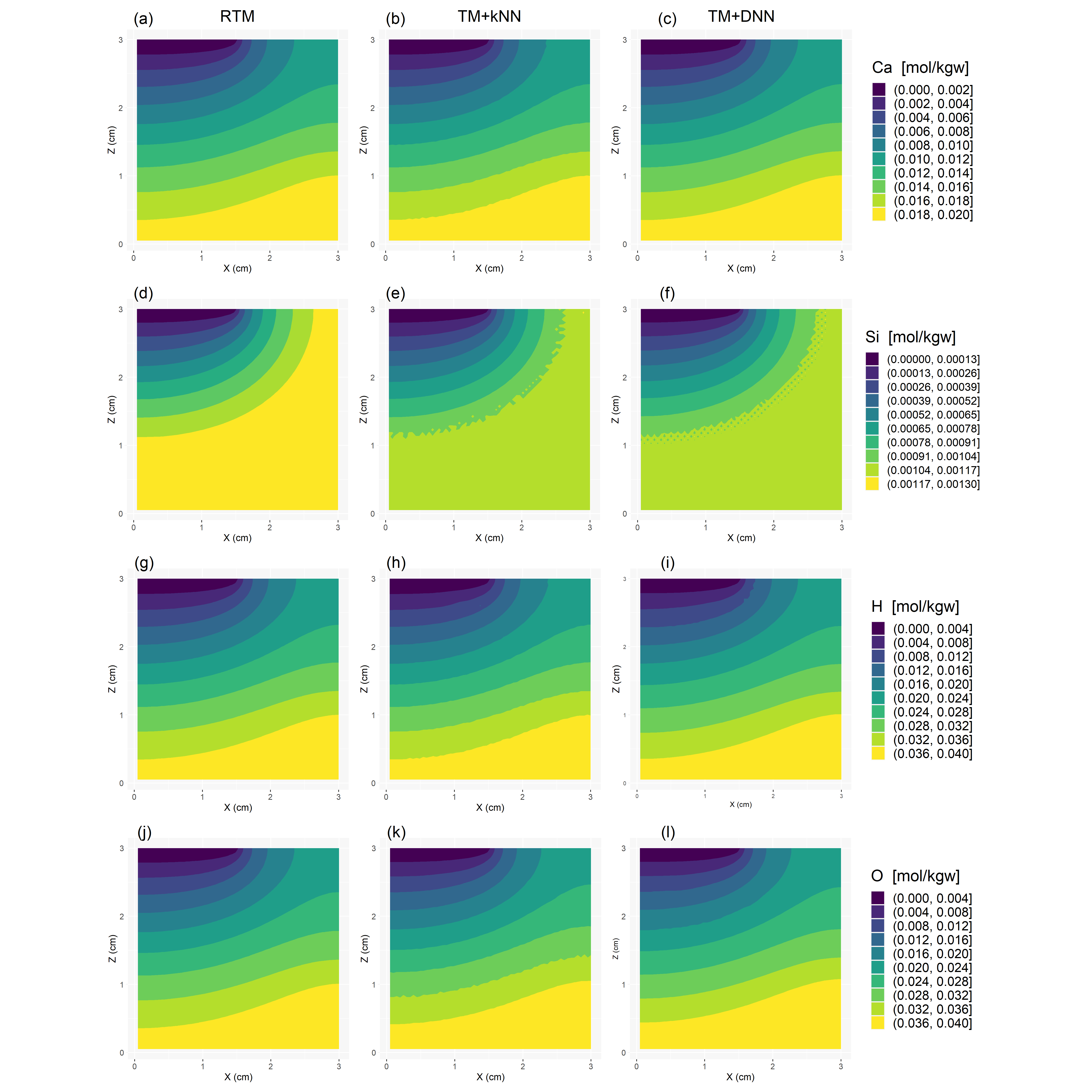}
	\caption{2D concentration profiles obtained for cement system 1 at the final time step of the 1-year simulation performed for the diffusion case. RTM means the original HP$_{\rm{4C}}$ model, TM+kNN denotes the Hydrus transport model coupled with our kNN geochemical emulator (HPx$_{\rm py}$-kNN) and TM+DNN signifies the Hydrus transport model coupled with our DNN geochemical emulator (HPx$_{\rm py}$-DNN). The first to fourth row present profiles for $Ca^{\rm{conc}}$, $Si^{\rm{conc}}$, $H^{\rm{conc}}$, and $O^{\rm{conc}}$, respectively. The considered grid size is 61 $\times$ 61.}
	\label{fig6}
\end{figure}

\begin{figure}[h!]
	\noindent\hspace{-1cm}\includegraphics[width=45pc]{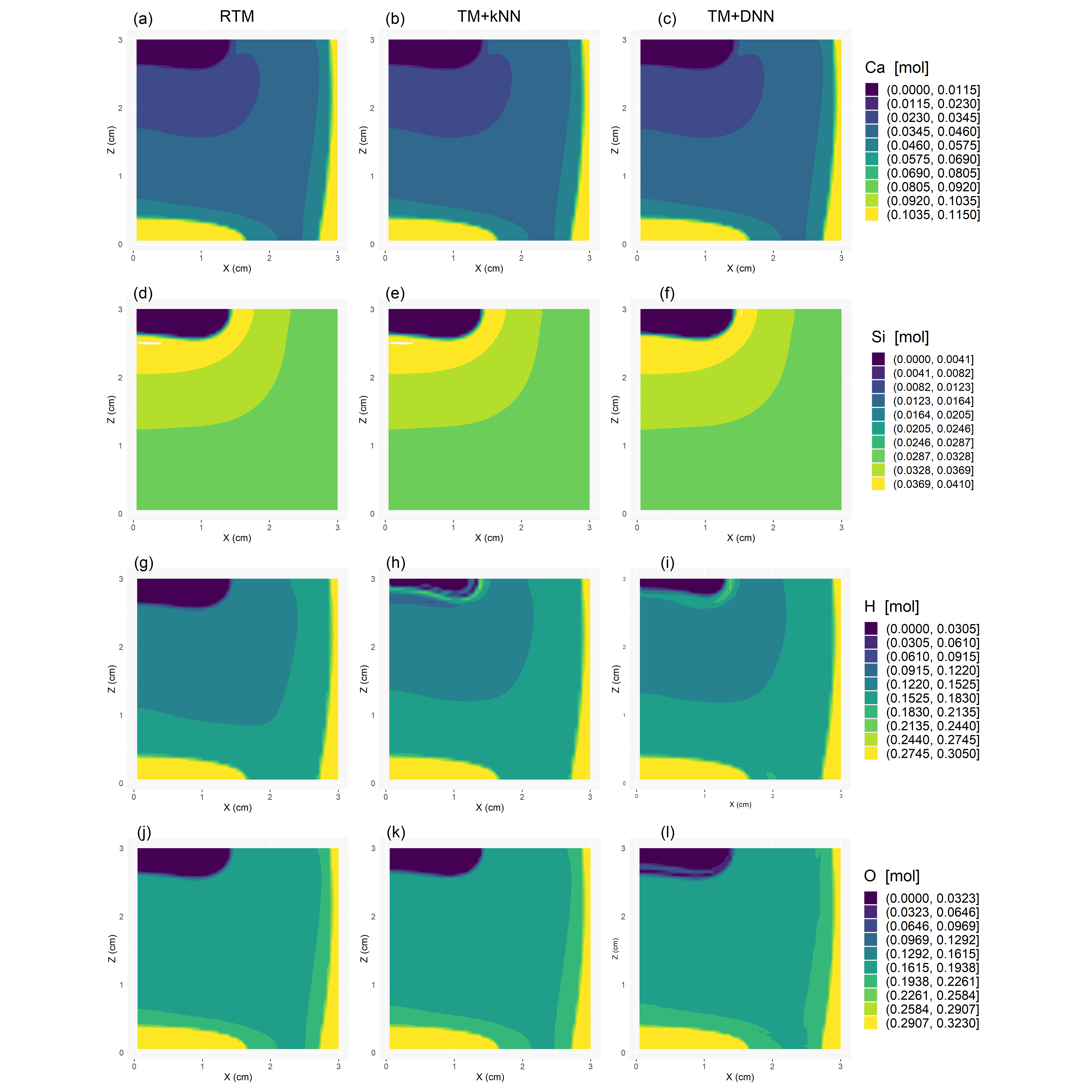}
	\caption{2D solid amount profiles obtained for cement system 1 at the end of the 2-year simulation performed for the advection-dispersion case. RTM means the original HP$_{\rm{4C}}$ model, TM+kNN denotes the Hydrus transport model coupled with our kNN geochemical emulator (HPx$_{\rm py}$-kNN) and TM+DNN signifies the Hydrus transport model coupled with our DNN geochemical emulator (HPx$_{\rm py}$-DNN). The first to fourth row present profiles for $Ca^{\rm{solid}}$, $Si^{\rm{solid}}$, $H^{\rm{solid}}$, and $O^{\rm{solid}}$, respectively. The considered grid size is 61 $\times$ 61. The results for the 121 $\times$ 121 grid are rather similar.}
	\label{fig7}
\end{figure}

\begin{figure}[h!]
	\noindent\hspace{-1cm}\includegraphics[width=45pc]{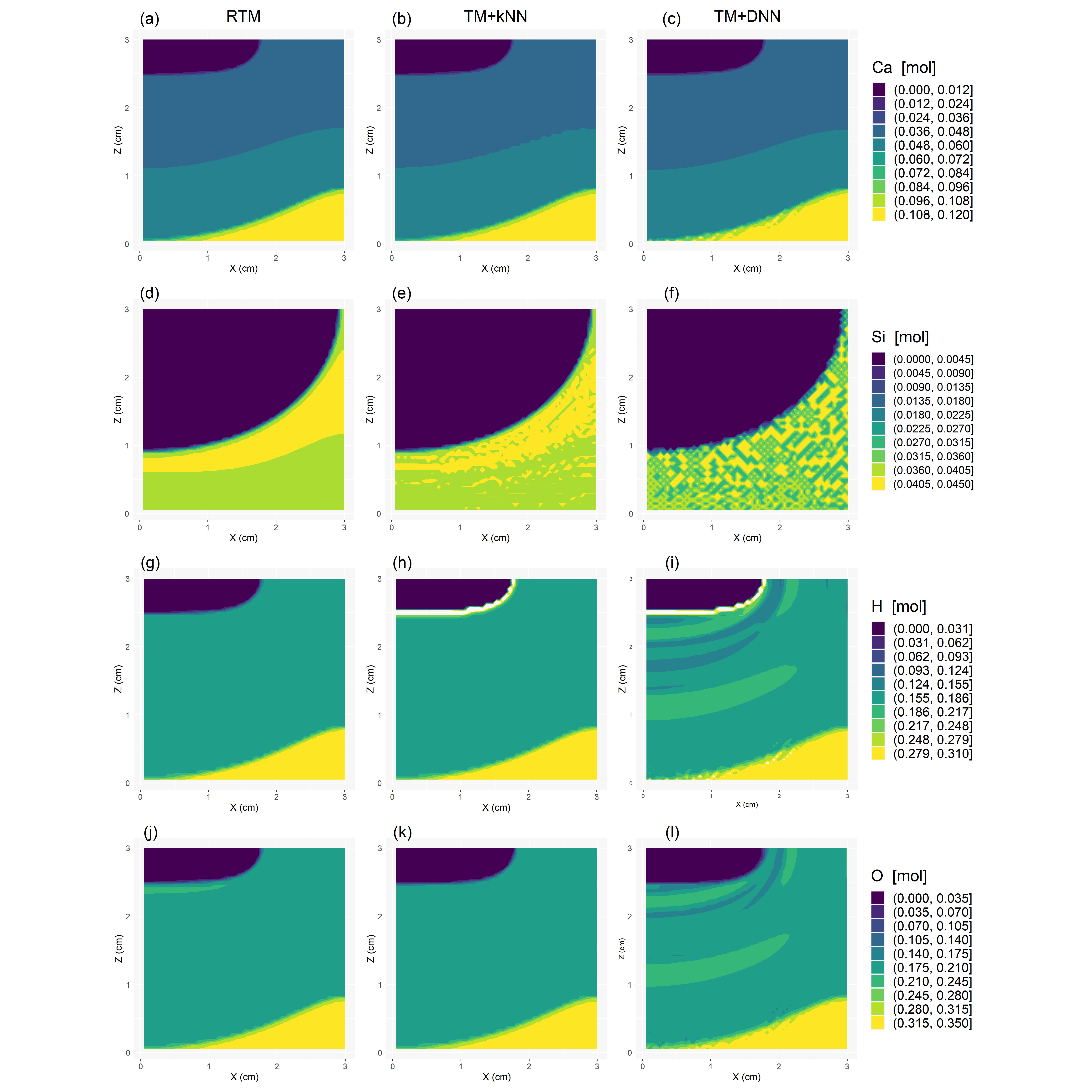}
	\caption{2D solid amount profiles obtained for cement system 1 at the final time step of the 1-year simulation performed for the diffusion case. RTM means the original HP$_{\rm{4C}}$ model, TM+kNN denotes the Hydrus transport model coupled with our kNN geochemical emulator (HPx$_{\rm py}$-kNN) and TM+DNN signifies the Hydrus transport model coupled with our DNN geochemical emulator (HPx$_{\rm py}$-DNN). The first to fourth row present profiles for $Ca^{\rm{solid}}$, $Si^{\rm{solid}}$, $H^{\rm{solid}}$, and $O^{\rm{solid}}$, respectively. The considered grid size is 61 $\times$ 61.}
	\label{fig8}
\end{figure}

This section focuses on reactive transport simulations with HPx$_{\rm py}$ within cement system 1, under both advective-dispersive and diffusive transport conditions. As written above, the domain sizes are both 61 $\times$ 61 and 121 $\times$ 121 for the advection-dispersion case and, because of computational constraints, solely 61 $\times$ 61  for the diffusion case. In addition, the simulation time period is 2 years for the advection-dispersion case and 1 year for the diffusion case. Figures \ref{fig3} and \ref{fig4} present times series of original and emulated Ca, Si, H and O concentrations at 5 locations within the 2D domain for advective-dispersive transport conditions, for both our kNN-based (HPx$_{\rm py}$-kNN) and DNN-based (HPx$_{\rm py}$-DNN) reactive transport codes. It is seen that HPx$_{\rm py}$-kNN and HPx$_{\rm py}$-DNN both induce a quite good simulation accuracy. Also, the results for the diffusive transport case are of similarly good quality (not shown). Figures \ref{fig5} - \ref{fig6} provide more insights into the HPx$_{\rm py}$-kNN and HPx$_{\rm py}$-DNN performances by displaying 2D Ca, Si, H and O concentration profiles at a given time. For each experiment and chemical component, this time is selected as to be well representative of the simulated dynamics. It is observed that the original and emulated images are visually almost indistinguishable for the advection-dispersion case (Figure \ref{fig5}). For the diffusion case, the emulators also perform quite well for Ca, H and O (Figures\ref{fig6}a - c, g - i and j - l), while some slight to moderate discrepancies appear at the concentration front for Si (Figures \ref{fig6}d - f). Notwithstanding, the Si concentration remains globally well predicted. Furthermore, Figures \ref{fig7} - \ref{fig8} present the original and emulated 2D solid amount profiles corresponding to Figures \ref{fig5} - \ref{fig6}. The original solid amount profiles are overall well approximated by HPx$_{\rm py}$-kNN and HPx$_{\rm py}$-DNN for the advection-dispersion case (Figure \ref{fig7}), even though some mismatch appears at the border of the fully depleted zone for the H component. As of the diffusion case (Figure \ref{fig8}), the same kind of mismatch is observed for the emulated solid amounts of H by HPx$_{\rm py}$-kNN while the emulated profiles by HPx$_{\rm py}$-DNN show some more discrepancies. Though we decided to present raw emulation results, we would like to stress that some if not all of the observed artifacts could likely be smoothed out by using some post-filtering such as median filtering.

The speedups associated with the considered problem are detailed in Table \ref{table3}. It is noted that the GPU-based DNN emulator allows for a speedup that is close to optimal. Indeed, The DNN speedups overall represent 85 \% to 95 \% of the maximum possible speedups (that is, speedups that would be obtained if the geochemcical calculations would come at no cost at all). The speedups associated with single-threaded kNN remain substantial but only amount to 57 \% - 65 \% of the corresponding maximum speedups. As detailed in section \ref{train_res1}, the used kNN and DNN implementations are found to be respectively 300 and 4000 faster than single-threaded PHREEQC when predicting 10,000 points all one once for this geochemical system. Based on these numbers one could have expected the achieved speedups to represent say 90 \% (kNN) or 99 \% (DNN) of the maximum possible ones. A large part of the gaps between achieved and maximum possible speedups is thus likely caused by the time required for communicating and exchanging data between the main C/C++ code and the Python-based emulators.

\begin{table}[h!]
	\caption{Speedups offered by the KNN and DNN emulators in HPx$_{\rm py}$ for the reactive transport simulations considered for cement system 1. The HPx$_{\rm{4C}}$ calculations involve the parallelization of PHREEQC over our 4 CPUs. The HPx$_{\rm{1C}}$ calculations are performed on a single CPU. The kNN predictions are performed on a single CPU using the scikit-learn implementation while the DNN predictions make use of our GPU. ML signifies the used machine learning method for emulation, TC denotes transport conditions (ADV: advection-dispersion, DIF: diffusion) and GS is the grid size. The maximum possible speedups associated with HPx$_{\rm{4C}}$ and HPx$_{\rm{1C}}$,  Max SP HPx$_{\rm{4C}}$ and  Max SP HPx$_{\rm{1C}}$, correspond to an hypothetical situation where the geochemical calculations incur zero computational cost.}
	\begin{center}
		\begin{tabular}{cccccccc}%
			\hline
			& & & & & & & \\
			ML & TC & GS & HPx$_{\rm{4C}}$ time (s) & SP HPx$_{\rm{4C}}$ & Max SP HPx$_{\rm{4C}}$ & SP HPx$_{\rm{1C}}$ & Max SP HPx$_{\rm{1C}}$ \\
			DNN & ADV & 61 $\times$ 61 & 3189 & 6.8 & 7.7 & 24.5 & 28.5 \\
			kNN & ADV & 61 $\times$ 61 & 3189 & 5.0 & 7.7 & 18.0 & 28.5 \\
			DNN & ADV & 121 $\times$ 121 & 23,337 & 5.0 & 5.2 & 17.0 & 18.0 \\
			kNN & ADV & 121 $\times$ 121 & 23,337 & 3.4 & 5.2 & 11.6 & 18.0 \\
			DNN & DIF & 61 $\times$ 61 & 25,448 & 4.2 & 4.9 & 16.2 & 19.1 \\
			kNN & DIF & 61 $\times$ 61 & 25,448 & 2.8 & 4.9 & 10.8 & 19.1 \\
			\hline
		\end{tabular}
	\end{center}
	\label{table3}
\end{table}

\FloatBarrier

\subsection{Al-C-Ca-S-Si Problem}

For this second cement system, the emulation problem consists of predicting at each time step of the RT simulation the (output) Al, C, Ca, S, Si, H and O aqueous concentrations (mol/kgw) from the (input) total amounts of Al, C, Ca, S and Si (mol). Here we focus on advective-dispersive transport only while similarly as for cement system 1, the considered domain sizes are 61 $\times$ 61 and 121 $\times$ 121. Moreover, the simulation time period is set to 6 years. 

As mentioned earlier, for this higher-dimensional problem it is observed that the used scikit-learn kNN implementation becomes prohibitively slow compared to HPx$_{\rm{4C}}$. We found that to get a good emulation accuracy the kNN training base needs to contain 1,000,000 samples (or more). This training base's size together with a 5-dimensional search space leads to an HPx$_{\rm py}$-kNN reactive transport simulation time that is comparable to that of HPx$_{\rm{4C}}$. Therefore, for this second cement system we built a custom kNN regressor around another kNN implementation contained in the FAISS package \citep{faiss2017}. Our used FAISS variant allows for GPU computing and is much faster than scikit-learn for this cement system, but is slightly less accurate due to the use of an approximate rather than exact nearest neighbor search \citep[see][for details]{faiss2017}.

\subsubsection{Training the DNN}
\label{train_res2}

Building a good training set to perform a kNN search and learn the weights and biases of our DNN turned out to be a complicated task in this case. This because to make useful kNN predictions and/or learn an useful DNN, the training set must be sufficiently representative of the geochemical conditions encountered during the reactive transport simulation one wish to perform with HPx$_{\rm py}$-kNN and the trained HPx$_{\rm py}$-DNN. In contrast to cement system 1, creating the training set by sampling the $X$-space with a controlled randomness between predefined lower and upper bounds did not prove successful. We tried that strategy by drawing as much as 4,000,000 5-dimensional $\textbf{x}$ vectors from the $X$-space using a Sobol low-discrepancy sequence \citep[][]{Sobol1967, Joe-Kuo2003}. Such low-discrepancy sampling scheme covers the 5-dimensional hypercube more uniformly than LHS. Despite a good performance on the test set (not shown), the resulting DNN accuracy in reactive transport mode was never deemed satisfying. In other words, no satisfying ``global" or ``universal" DNN emulator could be devised for this cement system. This is probably caused by the fact for this problem, the input (5 total amounts) and output (7 aqueous concentrations) spaces are quite nonlinearly related and both cover 6 to 10 orders of magnitudes depending on the considered element. Therefore, we resorted to the alternative training strategy detailed below. The latter basically tries to grasp the complex correlations and higher-order dependencies that exist between the elements of $\textbf{x}$ (total amounts, input space) for a given reactive transport simulation setup, in order to produce a training set that honors these between-input relationships.

\begin{itemize}
	\item Perform a ``cheap" full reactive transport simulation under the transport conditions and geochemistry of interest and collect the resulting $\textbf{x}$ -$\textbf{y}$  pairs of examples (for the considered grid nodes and time steps). Computational demand controls what domain size and simulation time period can be used for this cheap calculation. We used a modest 16 $\times$ 16 domain and a simulation time period of 10 years. The associated HPx$_{\rm{4C}}$ runtimes is 180 s.
	
	\item Fit a kernel density estimator (KDE) with a Gaussian kernel to the collected $\textbf{x}$ vectors (encapsulated in the $\textbf{X}$ array) and generate a fixed number of new input vectors, $\textbf{x}_{KDE}$. Then run PHREEQC for the $\textbf{X}_{KDE}$ set to get the corresponding output set, $\textbf{Y}_{KDE}$. Now apply the correction for porosity described in section \ref{train_res1} to the $\textbf{X}_{KDE}$ set and form the training set by merging the ensemble of $\textbf{x}$-$\textbf{y}$ pairs with that of the $\textbf{x}_{KDE}$-$\textbf{y}_{KDE}$ pairs. The number of produced unique examples by the considered cheap HPx$_{\rm{4C}}$ simulations varied between 10,000 and 50,000. The KDE-based enrichment of this dataset was deemed necessary to provide more input variability, thereby avoiding overfitting of the trained DNN and improving the kNN accuracy, while still honoring the complex between-input relationships. The number of KDE-generated samples was set as to obtain a total training set size of 1,000,000 examples. A key component of the approach is the bandwith parameter of the Gaussian KDE kernel which controls how much the KDE-generated samples depart from the original ensemble. After limited trial and error, we fixed the kernel bandwith to 0.0025 for the considered case studies.
	
\end{itemize}

\begin{figure}[h!]
	\noindent\hspace{-1.5cm}\includegraphics[width=50pc]{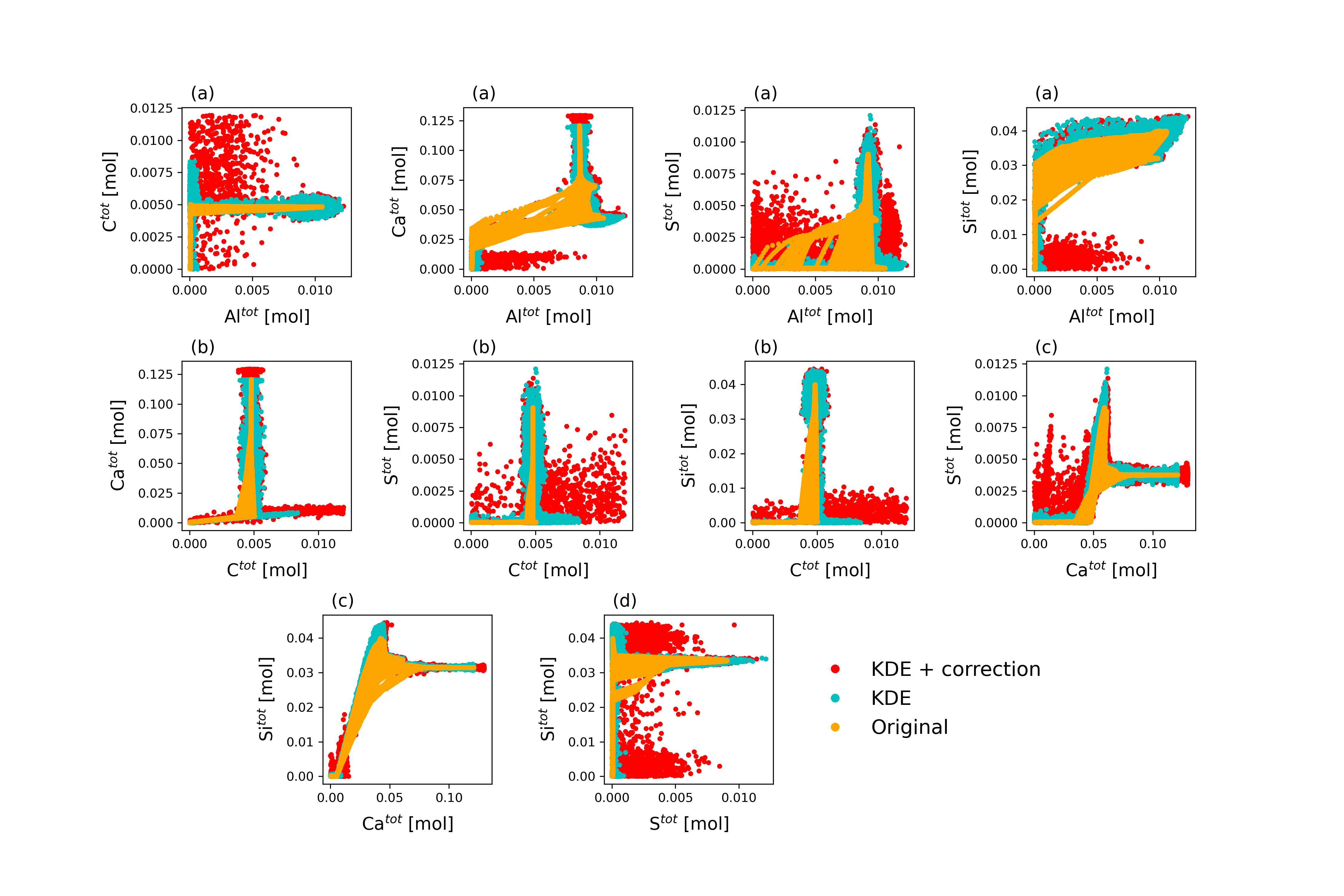}
	\caption{Scatter plots of the complex relationships between the five considered inputs for (1) the computationally cheap RT simulation of cement system 2 (orange dots) performed with the original HPx code, (2) the corresponding sampled points by kernel density estimation (KDE) using the selected bandwith (turquoise dots) and, (3) the same sampled KDE points after porosity correction (red dots).}
	\label{fig9}
\end{figure}

The scatter plots in Figure \ref{fig9} illustrate our training set creation procedure. The orange dots depict the pairwise relationships observed between the 5 elements of $\textbf{x}$ in the cheap simulation. The red and cyan dots in Figure \ref{fig9} represent the KDE-generated samples with the selected bandwidth, before and after applying the correction for porosity, respectively. Training of the DNN is achieved using the ensemble of original and KDE-corrected input points. We refer to this kind of dataset as RT-based, since it is based on an albeit cheap, full RT simulation. Furthermore, we refer to the obtained DNN and kNN emulators as ``local" emulators, since as opposed to the emulators constructed for cement system 1, the current emulators are only valid for the input conditions encapsulated in the RT-based training set.

\begin{figure}[h!]
	\noindent\hspace{-1cm}\includegraphics[width=45pc]{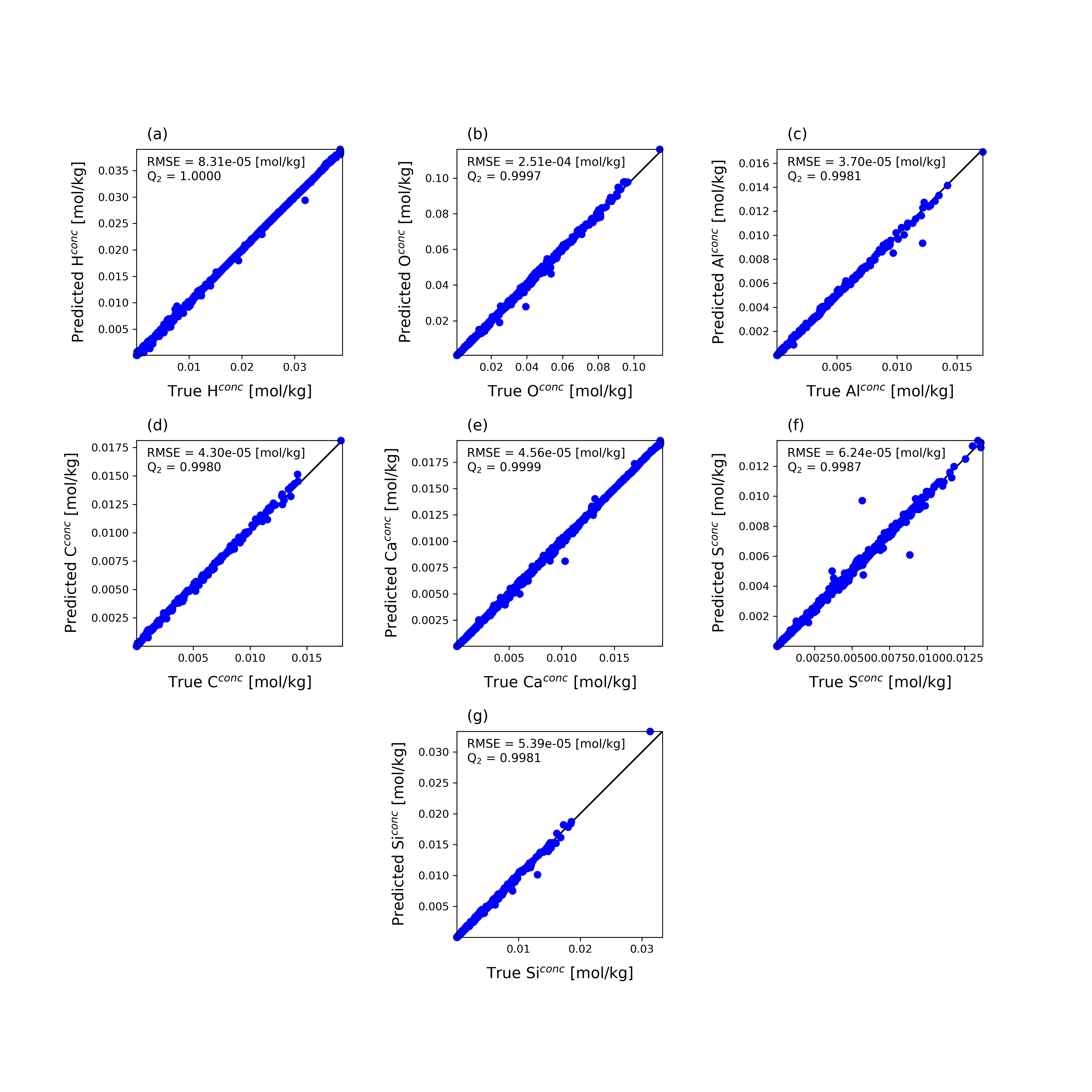}
	\caption{1-1 plots of local DNN emulation performance obtained for system 2 when the local DNN is trained using 1,000,000 samples. The $x$-axis and $y$-axis present the original and emulated 10,000 independent test data points, respectively. The RMSE and $Q_2$ coefficient denote the root-mean-square-error and coefficient of determination in testing mode, respectively, between the original and emulated 10,000 test data points.}
	\label{fig10}
\end{figure}

Training performance of the DNN emulator is presented in Figure \ref{fig10}. Similarly as for cement system 1, about 90\% of the available data was used for the actual training of the DNN while the remaining 10\% were used a validation set to contol overfitting. Lastly, performance is evaluated for both the trained DNN and kNN emulators using an independent test set of 10,000 examples. Overall, the accuracy of our ``local " DNN emulator for this RT-inspired dataset is rather large with $Q_2$ values always greater or equal than 0.998. Training performance of the corresponding local kNN emulator is equally good (not shown). Regarding speedup, for this problem single-threaded PHREEQC achieves about 210 geochemical calculations per second on our used Intel\textsuperscript{\textregistered} i7 CPU while the GPU-based DNN and kNN emulators are both about 3000 times faster when predicting the 10,000 test points at once .

\subsubsection{Reactive transport simulation}
\label{prob2res}
\begin{figure}[h!]
	\noindent\hspace{-1cm}\includegraphics[width=45pc]{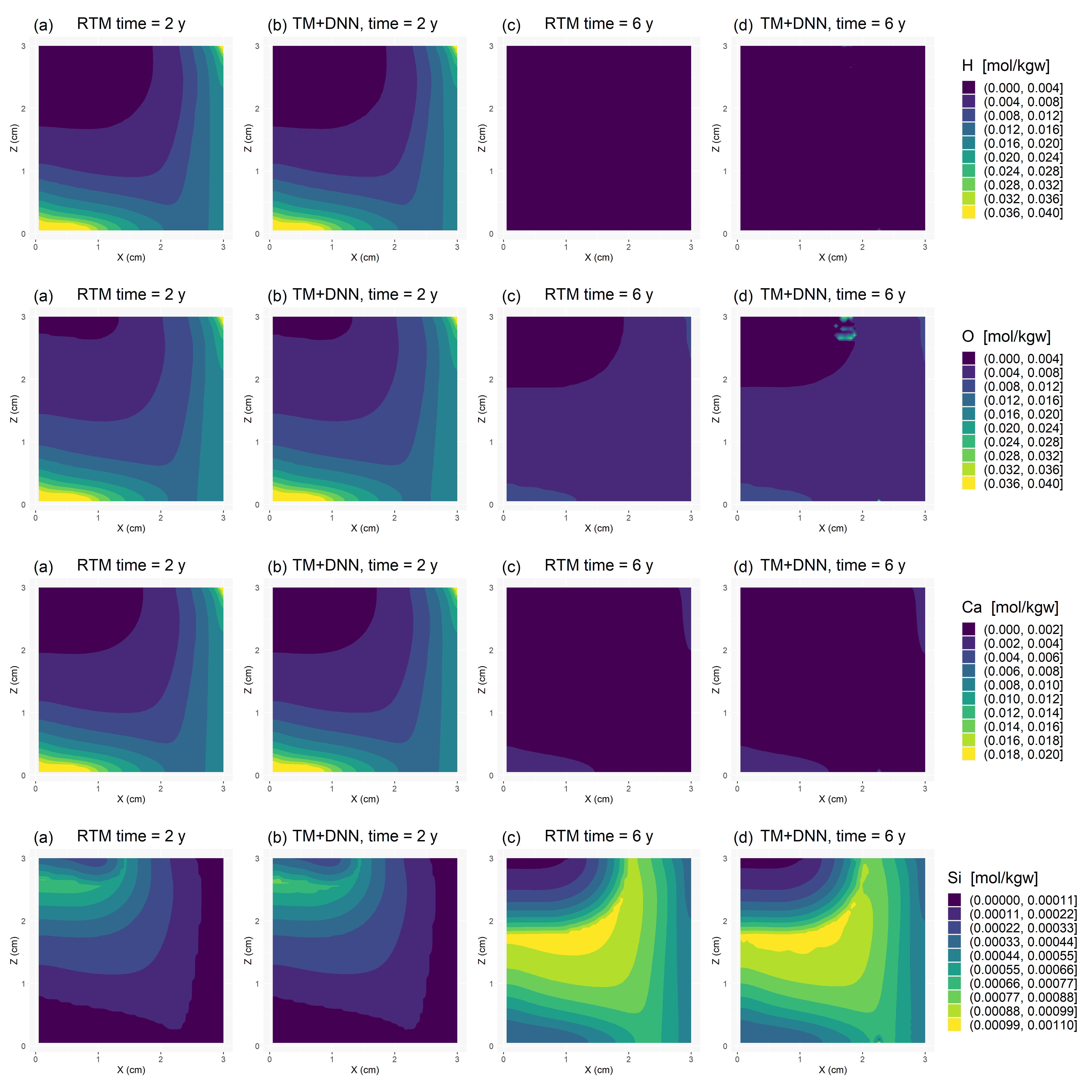}
	\caption{2D concentration profiles obtained for cement system 2 after 2 and 6 (final time step) years. RTM means the original HPx$_{\rm{4C}}$ model and TM+DNN denotes the Hydrus transport model coupled with our DNN geochemical emulator (HPx$_{\rm py}$-DNN). The first to fourth row present profiles for $Ca^{\rm{conc}}$, $Si^{\rm{conc}}$, $H^{\rm{conc}}$, and $O^{\rm{conc}}$, respectively. The considered grid size is 61 $\times$ 61.}
	\label{fig11}
\end{figure}

\begin{figure}[h!]
	\noindent\hspace{-1cm}\includegraphics[width=45pc]{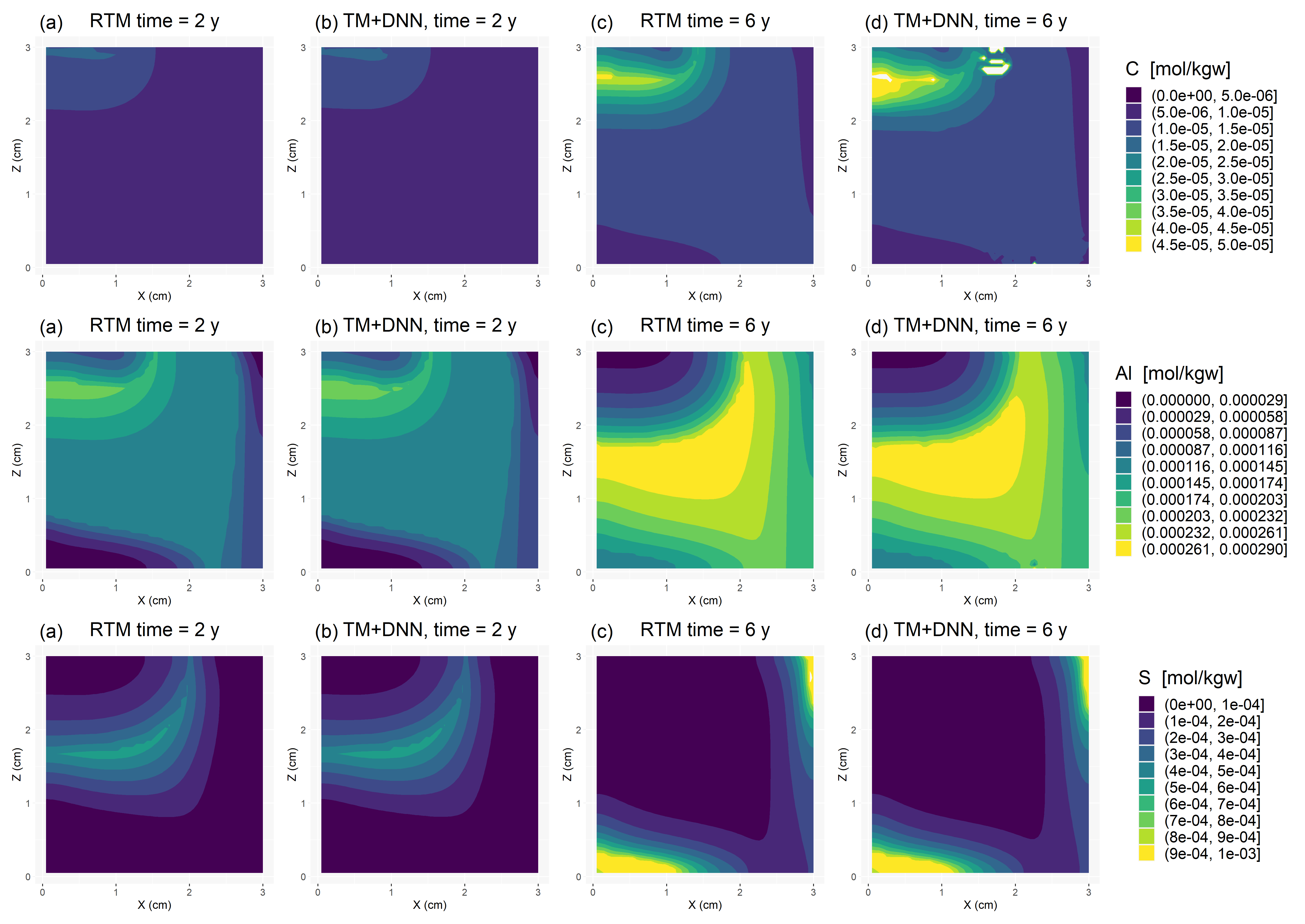}
	\caption{2D concentration profiles obtained for cement system 2 after 2 and 6 (final time step) years. RTM means the original HPx$_{\rm{4C}}$ model and TM+DNN denotes the Hydrus transport model coupled with our DNN geochemical emulator (HPx$_{\rm py}$-DNN). The first to third row present profiles for $C^{\rm{conc}}$, $Al^{\rm{conc}}$, and $S^{\rm{conc}}$, respectively. The considered grid size is 61 $\times$ 61.}
	\label{fig12}
\end{figure}

\begin{figure}[h!]
	\noindent\hspace{-1cm}\includegraphics[width=45pc]{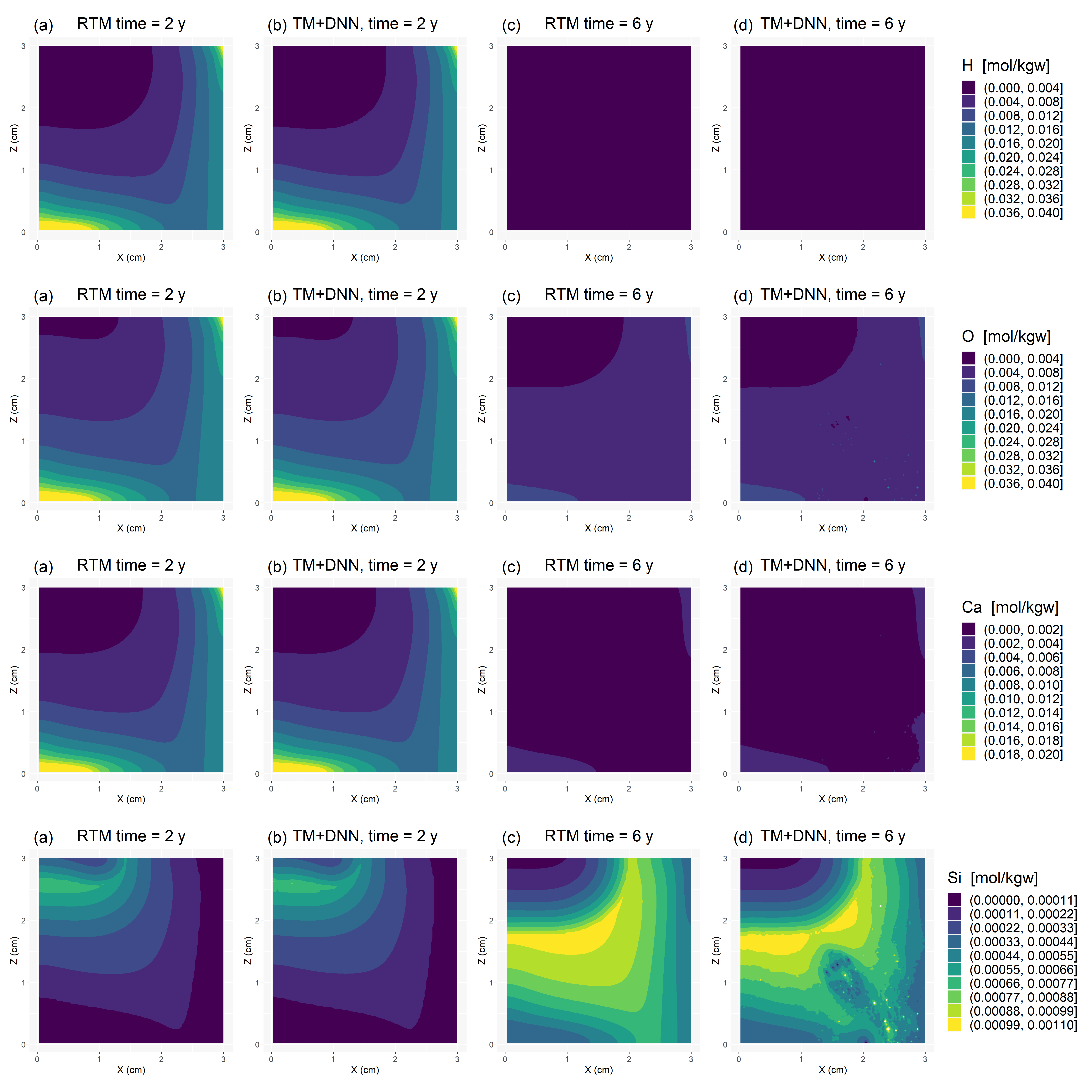}
	\caption{2D concentration profiles obtained for cement system 2 after 2 and 6 (final time step) years. RTM means the original HPx$_{\rm{4C}}$ model and TM+DNN denotes the Hydrus transport model coupled with our DNN geochemical emulator (HPx$_{\rm py}$-DNN). The first to fourth row present profiles for $Ca^{\rm{conc}}$, $Si^{\rm{conc}}$, $H^{\rm{conc}}$, and $O^{\rm{conc}}$, respectively. The considered grid size is 121 $\times$ 121.}
	\label{fig13}
\end{figure}

\begin{figure}[h!]
	\noindent\hspace{-1cm}\includegraphics[width=45pc]{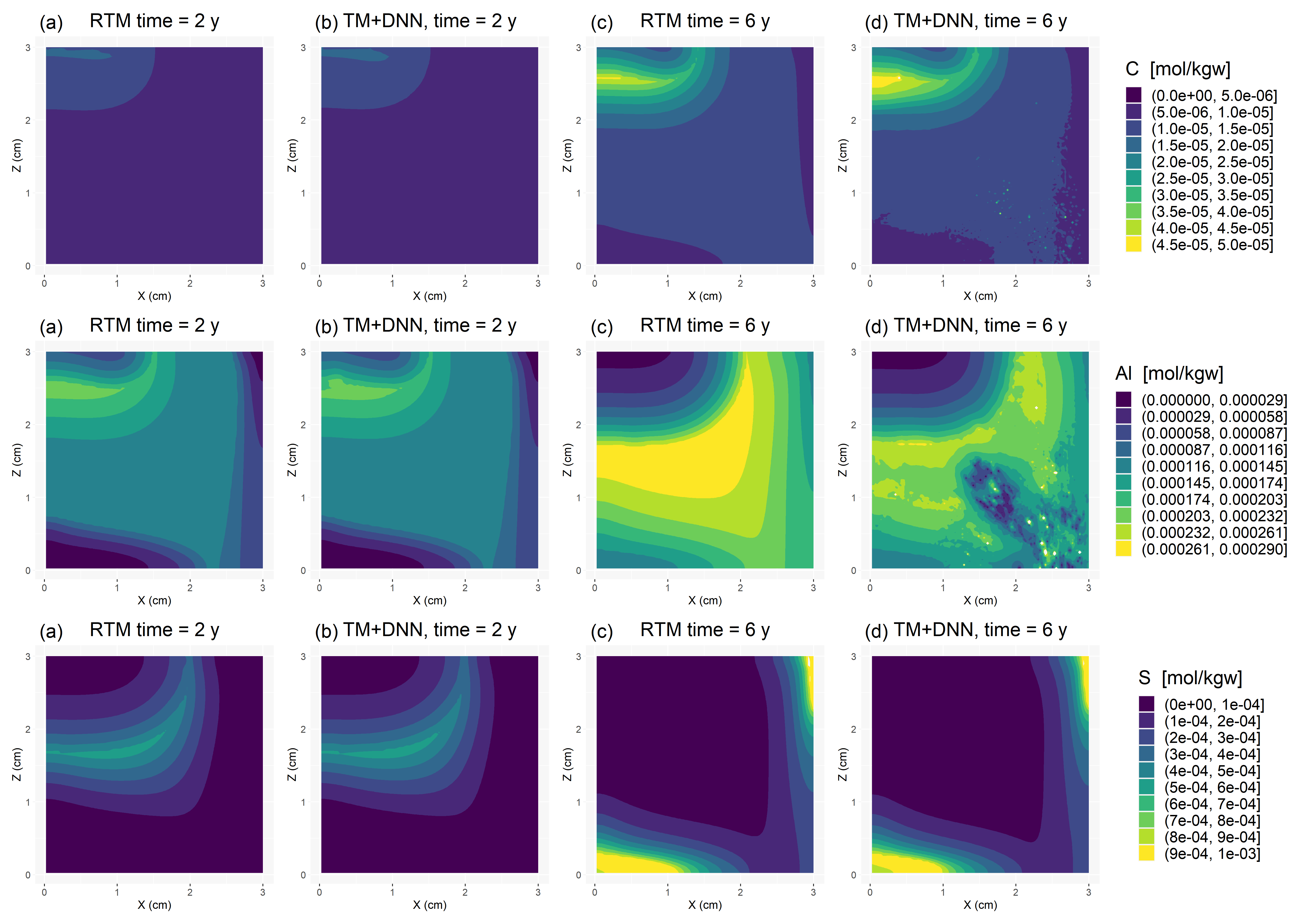}
	\caption{2D concentration profiles obtained for cement system 2 after 2 and 6 (final time step) years. RTM means the original HPx$_{\rm{4C}}$ model and TM+DNN denotes the Hydrus transport model coupled with our DNN geochemical emulator (HPx$_{\rm py}$-DNN). The first to third row present profiles for $C^{\rm{conc}}$, $Al^{\rm{conc}}$, and $S^{\rm{conc}}$, respectively. The considered grid size is 121 $\times$ 121.}
	\label{fig14}
\end{figure}

Our ``local" DNN performs rather well when applied to the 61 $\times$ 61 grid size and a time period of 6 years (Figures \ref{fig11} - \ref{fig12}). This for all components but C, for which some localized deviations appear between original and emulated 2D concentration profiles towards the end of the simulation period (Figure \ref{fig12}). When applied to the 121 $\times$ 121 grid, HPx$_{\rm py}$-DNN produces additional discrepancies for O, Si and Al towards the end of the simulation period (Figures \ref{fig13} - \ref{fig14}). Yet most of the observed artifacts could probably be smoothed out by using post-filtering. The associated speedups are listed in Table \ref{table4}. These speedups are larger to those obtained for cement system 1, with values between 8 and 9 when evaluated against HPx$_{\rm{4C}}$. These speedups represent about 85 \% to 90 \% of the maximum possible speedup (Table \ref{table3}). Overall, these findings indicate that for the considered problem, our RT-based training of a local DNN only works is the training set is sufficiently representative of the particular geochemical conditions encountered in the computationally demanding simulations, which is arguably not easy to achieve. This limitation is further discussed in section \ref{discussion}.

\begin{figure}[h!]
	\noindent\hspace{-1cm}\includegraphics[width=45pc]{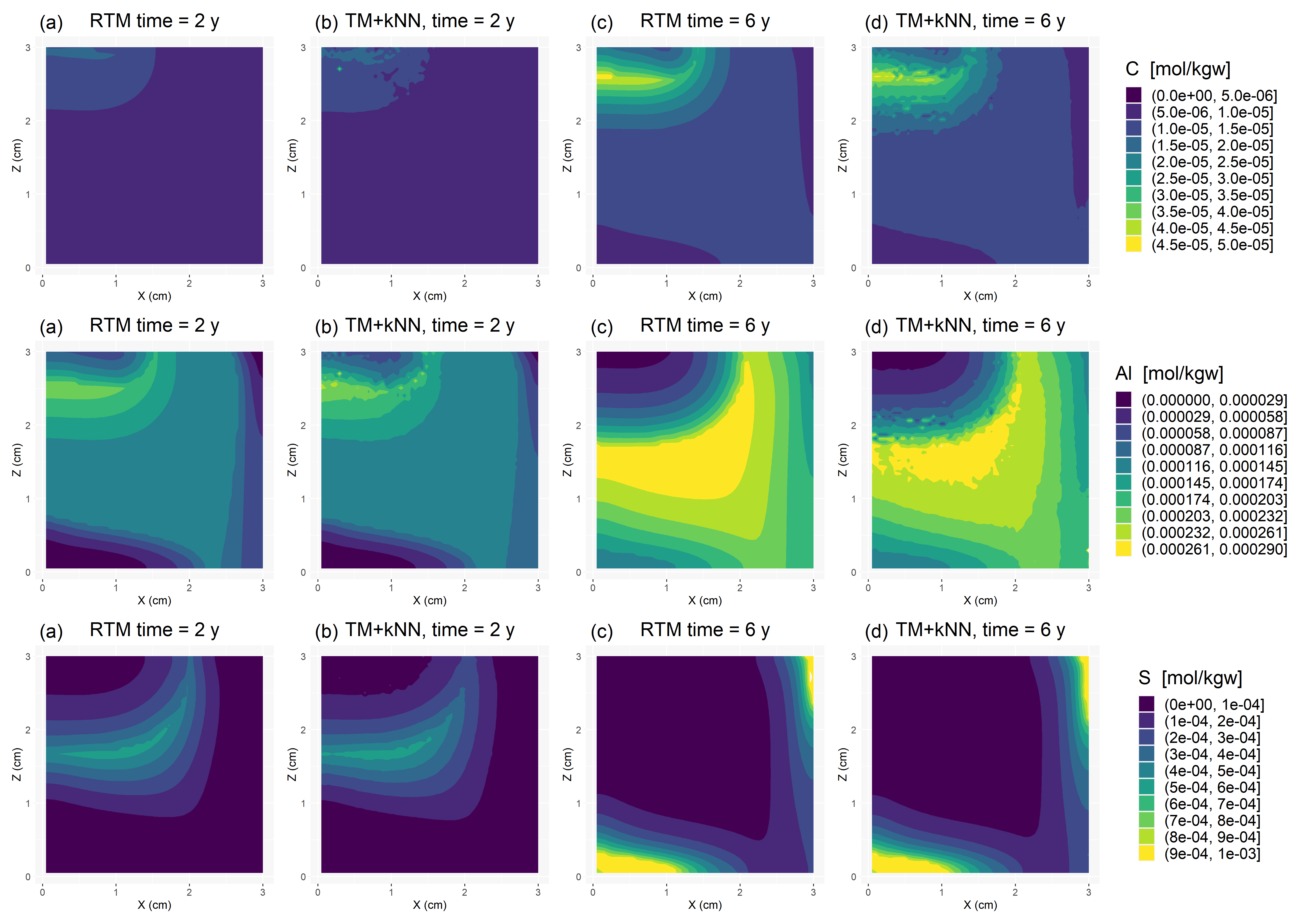}
	\caption{2D concentration profiles obtained for cement system 2 after 2 and 6 (final time step) years. RTM means the original HPx$_{\rm{4C}}$ model and TM+kNN denotes the Hydrus transport model coupled with our kNN geochemical emulator (HPx$_{\rm py}$-kNN).  The first to third row present profiles for $C^{\rm{conc}}$, $Al^{\rm{conc}}$, and $S^{\rm{conc}}$, respectively. The considered grid size is 61 $\times$ 61.}
	\label{fig15}
\end{figure}

\begin{figure}[h!]
	\noindent\hspace{-1cm}\includegraphics[width=45pc]{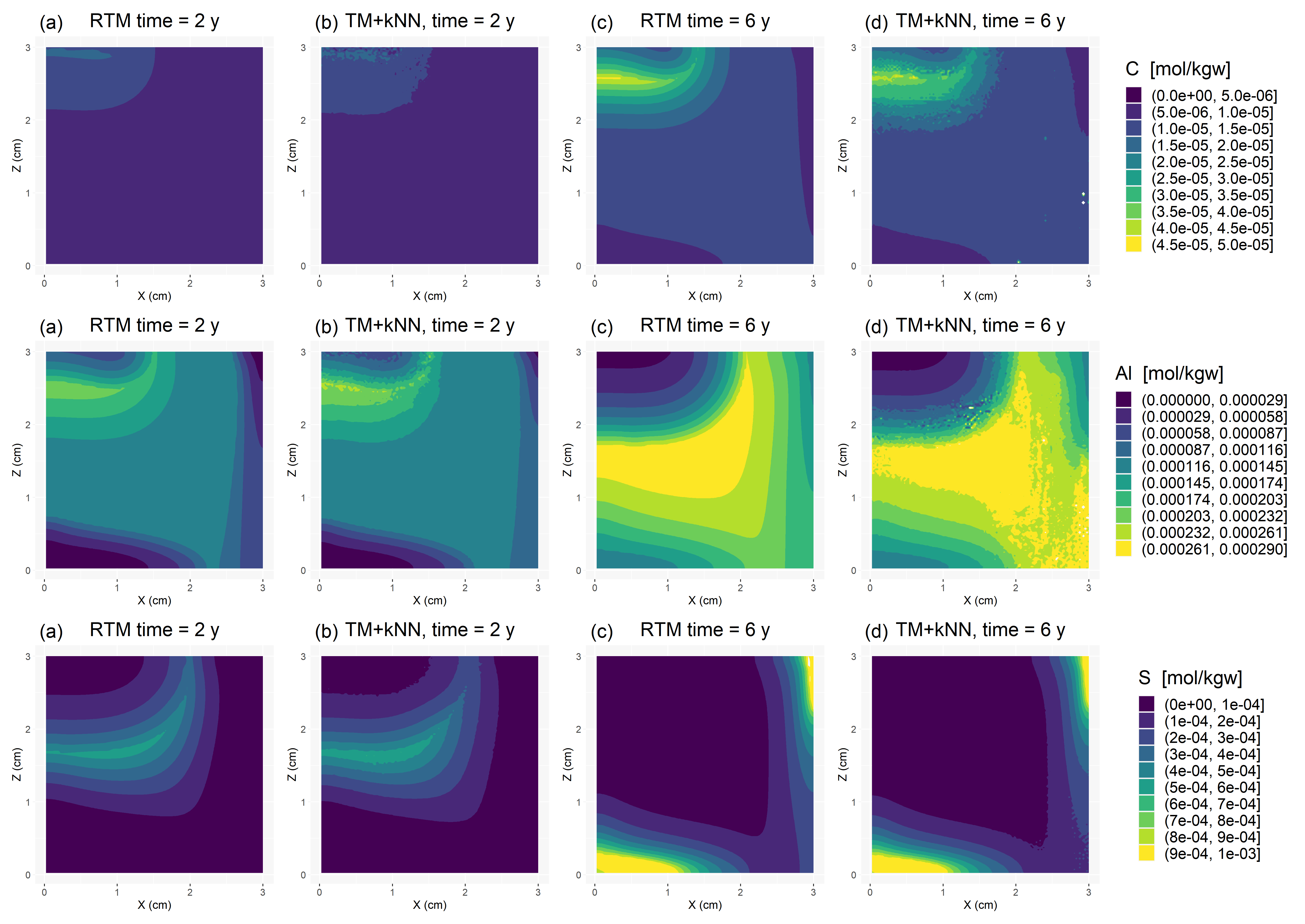}
	\caption{2D concentration profiles obtained for cement system 2 after 2 and 6 (final time step) years. RTM means the original HPx$_{\rm{4C}}$ model and TM+kNN denotes the Hydrus transport model coupled with our kNN geochemical emulator (HPx$_{\rm py}$-kNN). The first to third row present profiles for $C^{\rm{conc}}$, $Al^{\rm{conc}}$, and $S^{\rm{conc}}$, respectively. The considered grid size is 121 $\times$ 121.}
	\label{fig16}
\end{figure}

We note a more uniform behavior for HPx$_{\rm py}$-kNN across grid sizes than for HPx$_{\rm py}$-DNN. Here the results for the 121  $\times$ 121 grid are only slightly less accurate than those associated with the 61 $ \times$ 61 grid (see Figures \ref{fig15} - \ref{fig16} where for brevity we only show concentration profiles for the C Al and S components). Furthermore, whenever observed the discrepancies between original and emulated profiles are more regularly scattered than for HPx$_{\rm py}$-DNN. Note also that herein too, post-filtering could likely smooth out a large part of these deviations. In addition, owing to the use of a GPU to achieve the kNN calculations, the speedup provided by HPx$_{\rm py}$-kNN are as large as those provided by HPx$_{\rm py}$-DNN (Table \ref{table4}).

\begin{figure}[h!]
	\noindent\hspace{-1cm}\includegraphics[width=45pc]{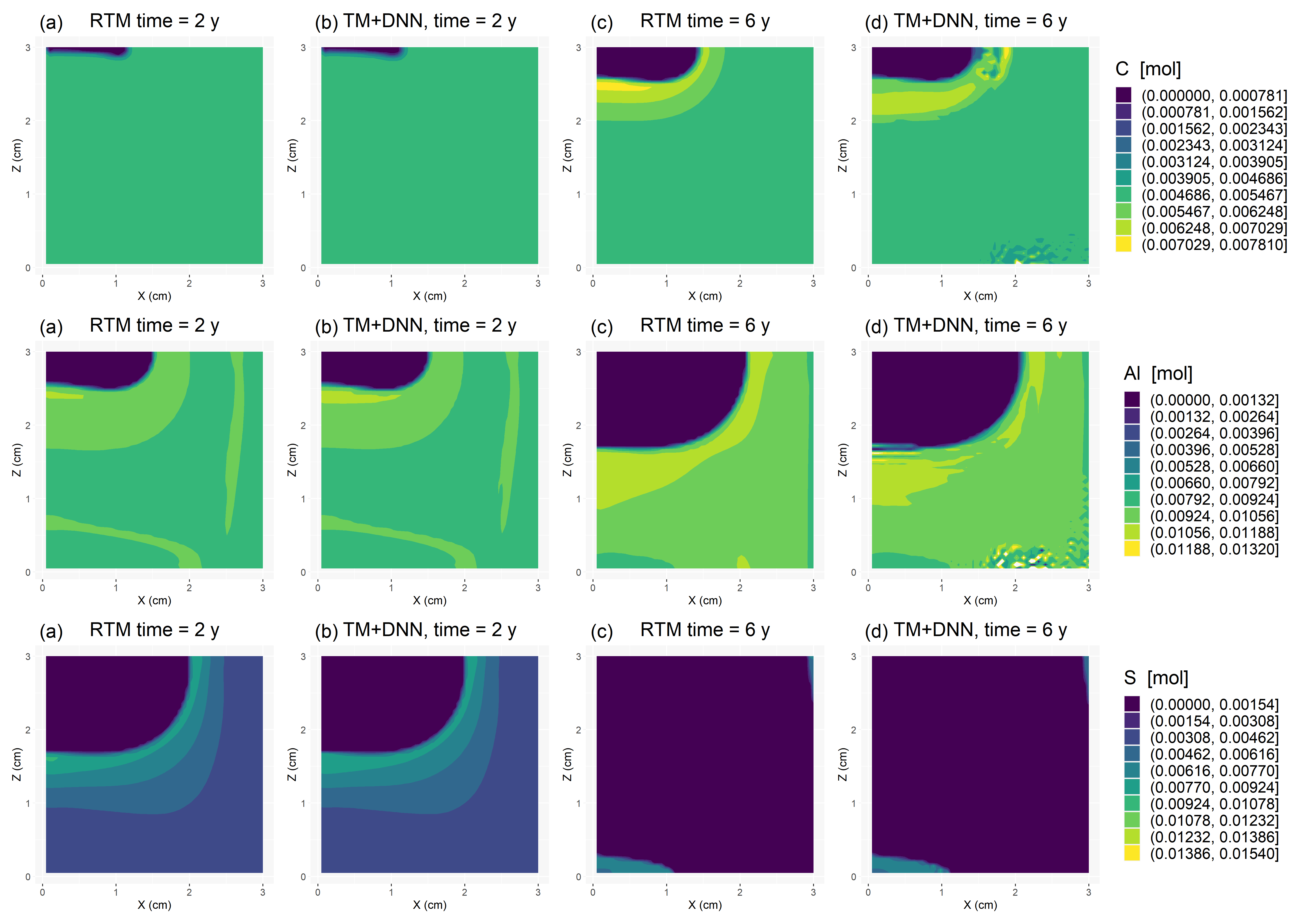}
	\caption{2D solid amount profiles of the C, Al and S chemical components of cement system 2 after 2 and 6 (final time step) years. RTM means the original HPx$_{\rm{4C}}$ model and TM+DNN denotes the Hydrus transport model coupled with our DNN geochemical emulator (HPx$_{\rm py}$-DNN). The first to third row present profiles for $C^{\rm{solid}}$, $Al^{\rm{solid}}$, and $S^{\rm{solid}}$, respectively The considered grid size is 61 $\times$ 61.}
	\label{fig17}
\end{figure}

\begin{figure}[h!]
	\noindent\hspace{-1cm}\includegraphics[width=45pc]{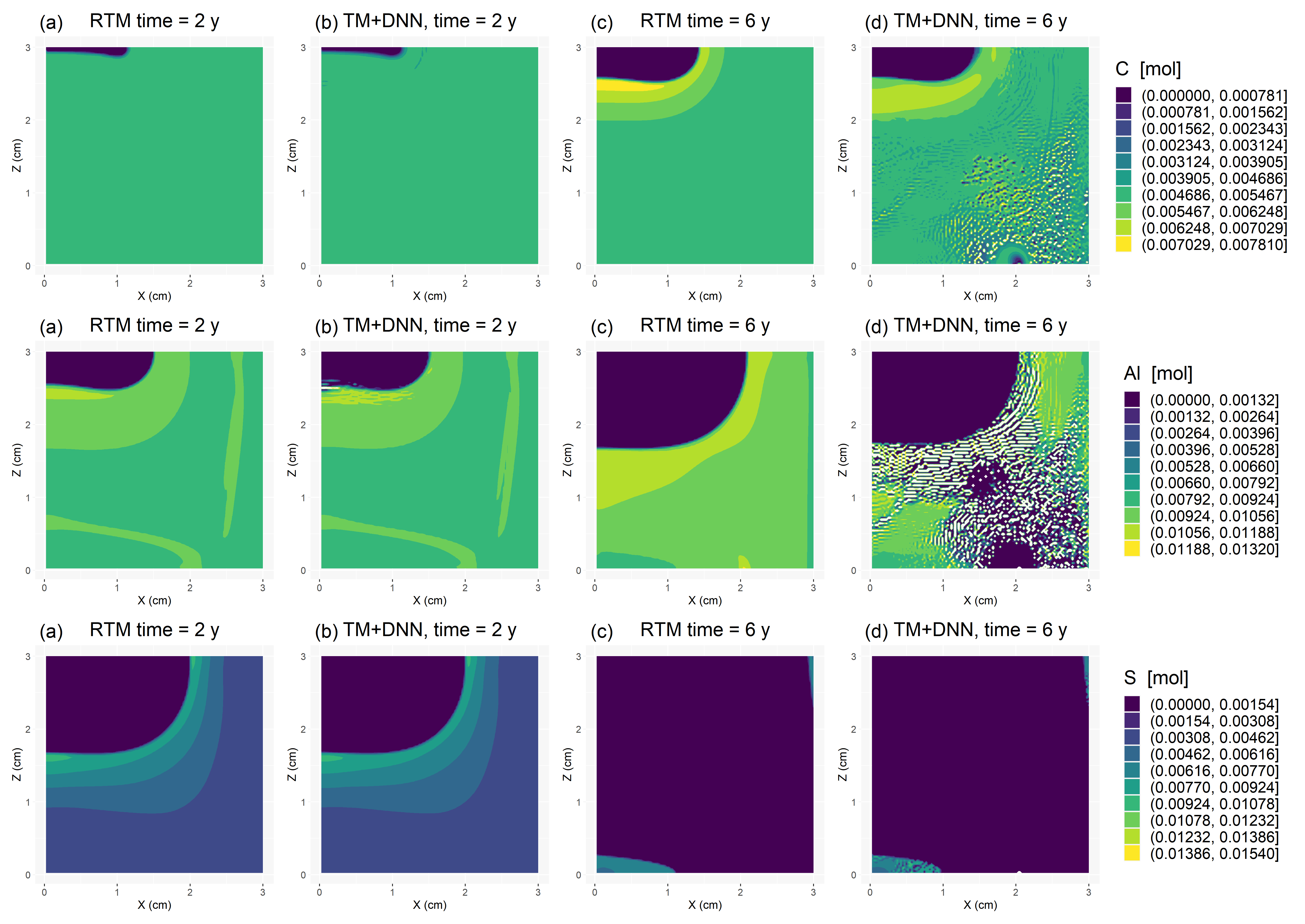}
	\caption{2D solid amount profiles of the C, Al and S chemical components of cement system 2 after 2 and 6 (final time step) years. RTM means the original HPx$_{\rm{4C}}$ model and TM+DNN denotes the Hydrus transport model coupled with our DNN geochemical emulator (HPx$_{\rm py}$-DNN). The first to third row present profiles for $C^{\rm{solid}}$, $Al^{\rm{solid}}$, and $S^{\rm{solid}}$, respectively The considered grid size is 121 $\times$ 121.}
	\label{fig18}
\end{figure}

\begin{figure}[h!]
	\noindent\hspace{-1cm}\includegraphics[width=45pc]{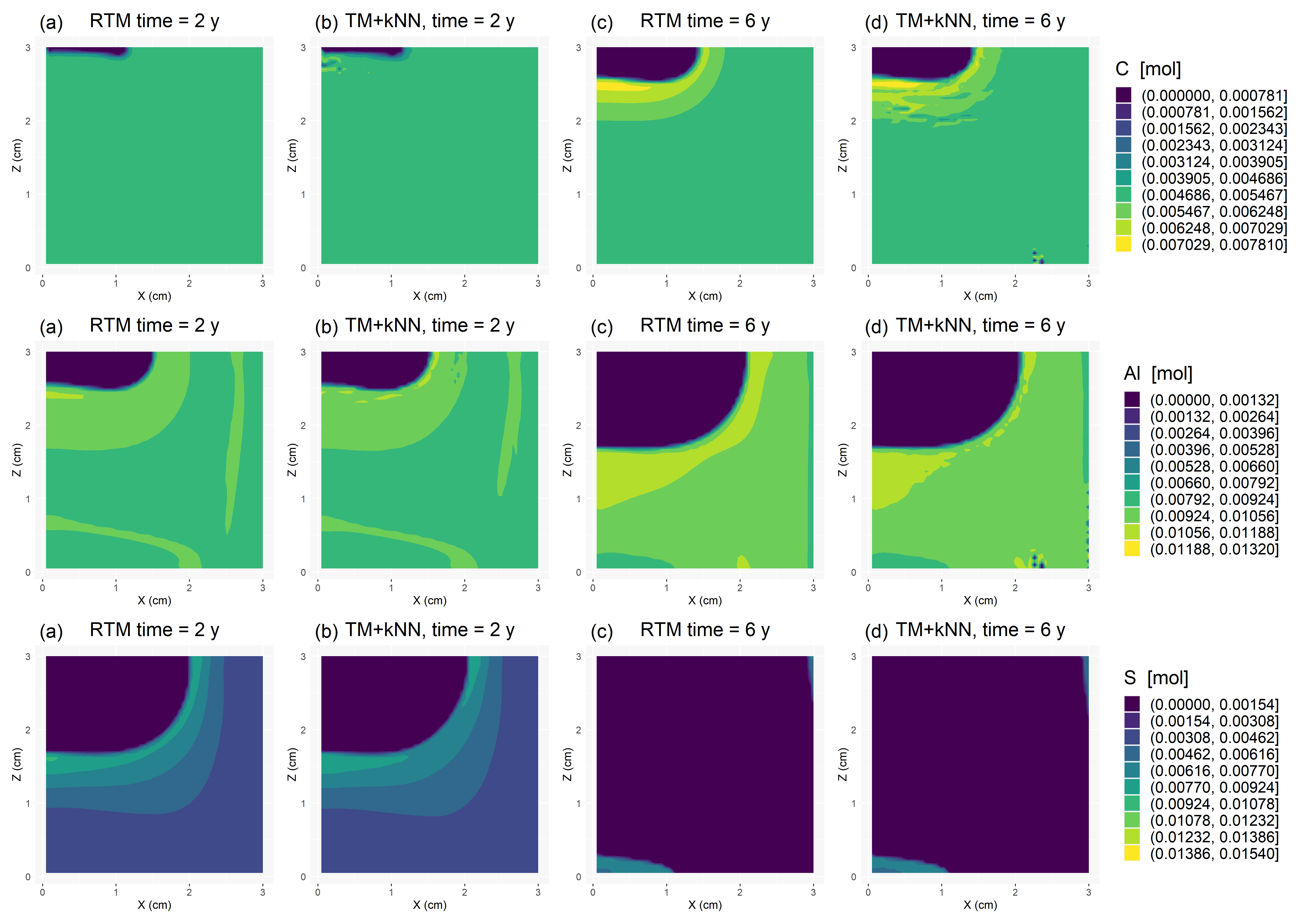}
	\caption{2D solid amount profiles of the C, Al and S chemical components of cement system 2 after 2 and 6 (final time step) years. RTM means the original HPx$_{\rm{4C}}$ model and TM+kNN denotes the Hydrus transport model coupled with our kNN geochemical emulator (HPx$_{\rm py}$-kNN). The first to third row present profiles for $C^{\rm{solid}}$, $Al^{\rm{solid}}$, and $S^{\rm{solid}}$, respectively The considered grid size is 61 $\times$ 61.}
	\label{fig19}
\end{figure}

\begin{figure}[h!]
	\noindent\hspace{-1cm}\includegraphics[width=45pc]{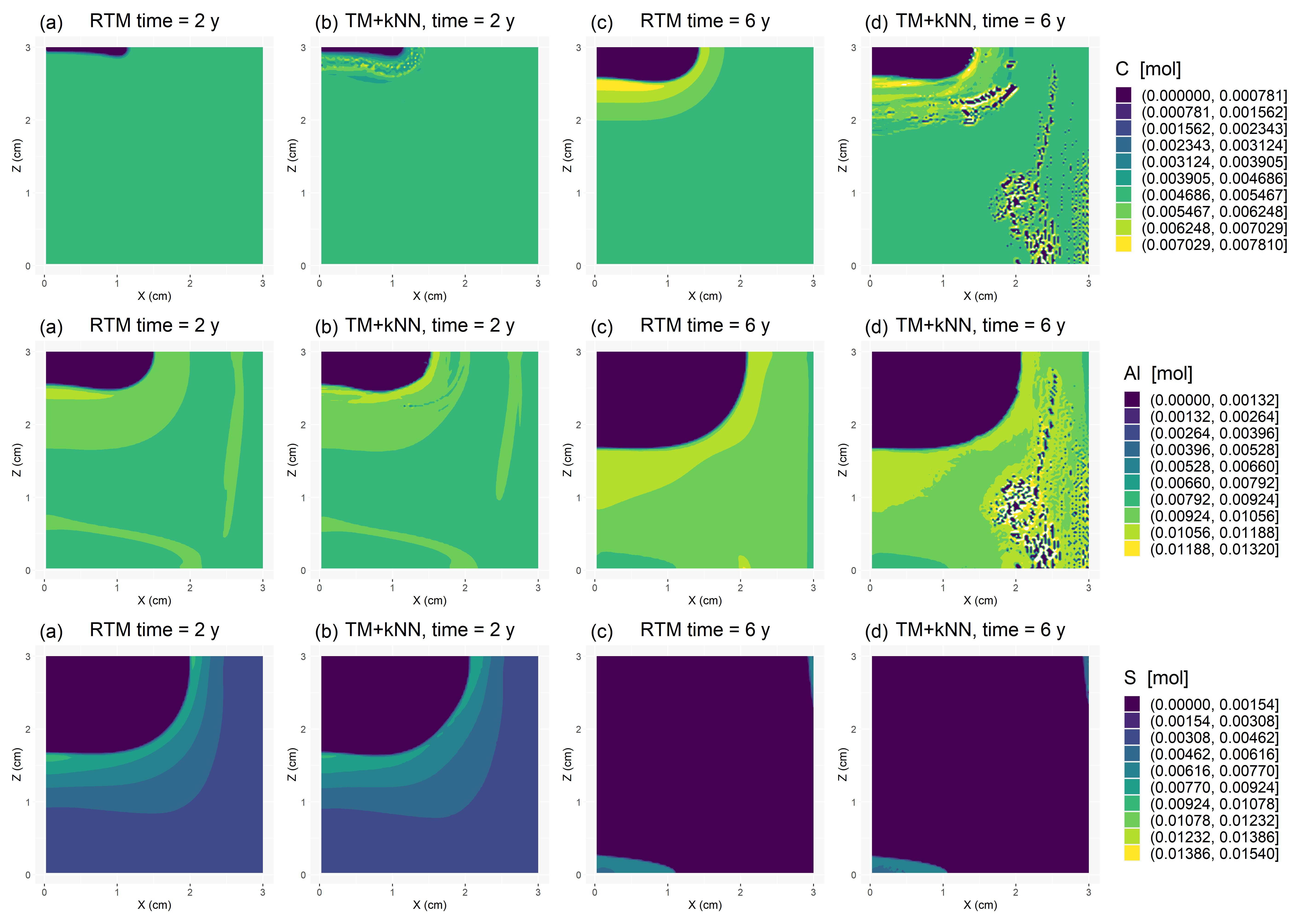}
	\caption{2D solid amount profiles of the C, Al and S chemical components of cement system 2 after 2 and 6 (final time step) years. RTM means the original HPx$_{\rm{4C}}$ model and TM+kNN denotes the Hydrus transport model coupled with our kNN geochemical emulator (HPx$_{\rm py}$-kNN). The first to third row present profiles for $C^{\rm{solid}}$, $Al^{\rm{solid}}$, and $S^{\rm{solid}}$, respectively The considered grid size is 121 $\times$ 121.}
	\label{fig20}
\end{figure}

With respect to the emulated solid amounts, the HPx$_{\rm py}$-DNN results look visually good for the the 61 $\times$ 61 grid. This is shown in Figure \ref{fig17} for the C, Al and S chemical components, while emulation of the H, O Ca and Si chemical components is globally of similar quality (not shown). Nevertheless, significant deviations appear towards the end of the simulation period for every chemical component (see Figure \ref{fig18} for the C, Al and S chemical components, emulation of the H, O Ca and Si chemical components shows the same level of mismatch). The HPx$_{\rm py}$-kNN predictions are also fairly accurate for the 61 $\times$ 61 (see Figure \ref{fig19} for the C, Al and S chemical components, emulation of the H, O Ca and Si chemical components exhibits a globally similar quality) while some discrepancies show up at the end of the simulation (Figure \ref{fig20}). However, the mismatch is less pronounced than for HPx$_{\rm py}$-DNN. Overall, HPx$_{\rm py}$-kNN appears to be somewhat more robust than HPx$_{\rm py}$-DNN for this cement system, while providing the same (large) speedup.

\begin{table}[h!]
	\caption{Speedups offered by the KNN and DNN emulators in HPx$_{\rm py}$ for the reactive transport simulations considered for cement system 2. The HPx$_{\rm{4C}}$ calculations involve the parallelization of PHREEQC over our 4 CPUs. The HPx$_{\rm{1C}}$ calculations are performed on a single CPU. Both the kNN and DNN predictions make use of a GPU. ML signifies the used machine learning method for emulation, BC denotes the type of flow boundary conditions and GS is the grid size. The maximum possible speedups associated with HPx$_{\rm{4C}}$ and HP$_{\rm{1C}}$,  Max SP HPx$_{\rm{4C}}$ and  Max SP HPx$_{\rm{1C}}$, correspond to an hypothetical situation where the geochemical calculations incur zero computational cost.}
	\begin{center}
		\begin{tabular}{cccccccc}%
			\hline
			& & & & & & & \\
			ML & BC & GS & HPx$_{\rm{4C}}$ time (s) & SP HPx$_{\rm{4C}}$ & Max SP HPx$_{\rm{4C}}$ & SP HPx$_{\rm{1C}}$ & Max SP HPx$_{\rm{1C}}$ \\
			DNN & ADV & 61 $\times$ 61 & 21,415 & 8.2 & 9.0 & 30.3 & 33.1\\
			kNN & ADV & 61 $\times$ 61 & 21,415 & 7.9 & 9.0 & 28.9 & 33.1\\
			DNN & ADV & 121 $\times$ 121 & 199,841 & 8.2 & 9.5 & 29.9 & 32.8\\
			kNN & ADV & 121 $\times$ 121 & 199,841 & 8.5 & 9.5 & 31.3 & 32.8\\
			\hline
		\end{tabular}
	\end{center}
	\label{table4}
\end{table}

\FloatBarrier

\section{Discussion}
\label{discussion}

We have shown that for a Ca-Si-H-O cement system involving leaching from hardened cement paste with portlandite and a solid-solution representation of the C-S-H phase, standard DNNs and kNN algorithms can emulate quite accurately the geochemical solver of a RT code (HPx) while offering large speedups. When moving to a more complex cement system which comprises 7 components: Al-C-Ca-S-Si-H-O, we found that no emulator that is appropriately accurate over the full space of possible geochemical conditions can be devised. Therefore, we focused on the creation of local DNN and kNN emulators that are locally accurate over the specific geochemical conditions of the computationally demanding simulation of interest. Yet the associated predictive accuracy is no longer close to perfect. Instead deviations from the original concentration profiles appear for some components towards the end of the simulation. Overall, The GPU-powered HPx$_{\rm py}$-DNN and HPx$_{\rm py}$-kNN codes offer a speedup that is close to optimal. The following remarks are in order.

Our tests reveal that the accuracy of the local DNN emulator is highly sensitive to the quality or degree of representativeness of the dataset the DNN is trained with. The feasibility of our strategy based on running a cheap full RT simulation to create initial training points that are subsequently enriched by KDE-based sampling thus appears to be largely case-specific for HPx$_{\rm py}$-DNN. Regarding kNN, HPx$_{\rm py}$-kNN is found to be moderately more robust in this setting, while providing the same speedup as HPx$_{\rm py}$-DNN when the kNN search is achieved on a GPU. Yet relevant chemical conditions might also only appear with finer meshes and/or shorter time steps than those used with the cheap RT simulation. If our random exploration of the interesting areas defined by the cheap RT simulation does not capture enough key information, then the resulting DNN or kNN emulator will be inaccurate. How to choose the right settings in the cheap simulation is thus an open problem. 

It might be that a larger DNN, that is a DNN with either more layers or more neurons, or both, would allow for training a proper ``global" DNN, that is, a DNN that similarly as for our first cement system is valid over the full range of potential geochemistry. This will be investigated in future work. As always with DNNs, this task will require careful training and overfitting control before the DNN can possibly be used.

Note that kNN is sensitive to the so-called curse of dimensionality. As the dimensionality of the input space increases, for a given database size the euclidean distances between each pair of input points will become increasingly similar, thereby preventing kNN to find useful closest neighbors. In other words, there will be a point where each considered input vector will have many equally close neighbors that are geochemically very different. For a given dimensionality, the impact of the curse of dimensionality on the kNN search depends on both the inner data structure and the database size. The more random the data and the less the number of training examples, the more the search is affected. Whether our kNN emulator can still work in 10 to 20 dimensions thus remains to be tested. With respect to our used GPU implementation, which provides the necessary speed for our second case study, it is worth pointing out that GPUs have limited memory availability compared to CPUs. This is not at all a problem for the considered case studies but might become an issue for chemical systems with many more inputs and outputs and/or for which a larger training base is required.

Although it suffers from the curse of dimensionality, an advantage of the kNN technique is its simplicity. Since it does not require offline training, a training base could be built on the fly during a full RT simulation and then subsequently used for emulation when deemed sufficiently informative. Or if full geochemical calculations are performed within an accept/reject framework of the emulated predictions, the kNN training base can be directly enriched with the calculated points. This is in essence similar to the on-demand surrogate modeling method described in \citet{Leal2017} and \citet{Leal2020} where an efficient clustering strategy is used instead of a nearest neighbor search.

As stated earlier, \citet{Leal2020} and \citet{Delucia-Kuhn2021} propose to replace the emulated geochemical prediction by a full calculation online during the RT simulation, if a quality criterion is not satisfied. Implementing this interesting possibility within our framework would be straightforward. Yet the definition of this criterion is not easy. The criterion by \citet{Leal2020} is based upon acceptable changes in chemical potential, which for a chemical solver relying on the law of mass actions (as PHREEQC in HPx) could possivbly be replaced by the activity or apparent chemical potential as outlined in \citet{Leal2016}. \citet{Delucia-Kuhn2021} resort to mass balance error. However, by design our implementation conserves the total mass before and after a reactive transport step in a given cell. More research is required to introduce an acceptance criterion for the emulated concentrations in our approach. 

\section{Conclusion}
\label{conclusion}

In this work, we investigate the potential of two machine learning approaches, deep neural networks (DNNs) and k-nearest neighbor (kNN) regression, to replace the geochemical solver in a reactive transport code, thereby providing a large speedup in reactive transport simulation. We focus on leaching from hardened cement paste involving either 4 or 7 components within 2D domains of sizes 61 $\times$ 61 and 121 $\times$ 121, and use the HPx reactive transport code as baseline. Our results show that after training, both our DNN-based and kNN-base codes, HPx$_{\rm py}$-DNN and HPx$_{\rm py}$-kNN, can predict geochemical outputs quite accurately for the simpler 4-components cement system while providing a 3 (single-threaded kNN) to 7 (GPU-based DNN) speedup compared to HPx with parallelized geochemical calculations over 4 cores. Benchmarking against single-threaded HPx, these speedups become 11 to 25. Things get more complicated for the more complex 7-components cement system. Here no emulator that is globally accurate over the full range of possible geochemical conditions could be built, neither with DNNs nor with kNN. Instead we therefore constructed ``local" emulators that are only valid over a relevant fraction of the input parameter space. This is done by running a computationally cheap full RT simulation to create a first set of training points. Case-specific computational demand controls what domain size and time period can be used. This initial training set is subsequently enriched by kernel density sampling to provide more input diversity while still honoring the complex between-input dependencies that are observed in the cheap RT simulation. The resulting training set is then either used to train the DNN emulator or serves as training base for the kNN emulator. This strategy is found to work but no longer (close to) perfectly, as both the HPx$_{\rm py}$-DNN and HPx$_{\rm py}$-kNN outputs present some discrepancies in the emulated series of 2D concentration (and associated solid amount) profiles compared to the baseline. Nevertheless, most of these irregularities can presumably be smoothed out to some extent using post-filtering. Also, the speedups associated with this 7-components problem are attractive: 8 to 9 (29 to 33) when evaluated against four-threaded (single-threaded) HPx and using a GPU for both the DNN and kNN regressions within HPx$_{\rm py}$-DNN and HPx$_{\rm py}$-kNN, respectively. Defining the maximum possible speedup as the computational gain in RT simulation that would be obtained if the geochemical calculations would be for free, we find that GPU-powered HPx$_{\rm py}$-DNN and HPx$_{\rm py}$-kNN achieve a close to optimal speedup. Future work will focus on how to improve accuracy for the considered 7-components cement system and other problems.

\section{Appendix: Details of the used DNN}
\label{appendix_a}

Let us denote the input and output space dimensions by $N_{\rm in}$ and $N_{\rm out}$, and a given reference number of hidden neurons as $NN_{\rm h}$. A fully connected layer (FC) with a selu activation function taking a $N_{\rm in}$-dimensional input and producing a $N_{\rm out}$-dimensional output using a self exponential activation function (SELU) is then referred to as $\left[FC_{\rm SELU}-I_{N_{\rm in}}-O_{N_{\rm out}}\right]$. We call $\left[FC_{\rm LIN}-I_{N_{\rm in}}-O_{N_{\rm out}}\right]$ the same layer but with a linear activation function.  From input to output layer, our DNNs are built as follows
\begin{itemize}
	\item $\left[FC_{\rm SELU}-I_{N_{\rm in}}-O_{N_{\rm h}}\right]$
	\item $\left[FC_{\rm SELU}-I_{N_{\rm h}}-O_{2N_{\rm h}}\right]$
	\item $\left[FC_{\rm SELU}-I_{2N_{\rm h}}-O_{4N_{\rm h}}\right]$
	\item $\left[FC_{\rm SELU}-I_{4N_{\rm h}}-O_{2N_{\rm h}}\right]$
	\item $\left[FC_{\rm SELU}-I_{2N_{\rm h}}-O_{N_{\rm h}}\right]$
	\item $\left[FC_{\rm LIN}-I_{N_{\rm h}}-O_{N_{\rm out}}\right]$
\end{itemize}
We set $N_{\rm h} = 32$ for the first cement system ($N_{\rm in} = 2$, $N_{\rm out} = 4$) and $N_{\rm h} = 128$ for the second cement system ($N_{\rm in} = 5$, $N_{\rm out} = 7$).

\bibliographystyle{unsrt}  


\end{document}